\documentclass[preprint,12pt]{elsarticle}

\usepackage{amssymb}
\usepackage{graphicx}
\usepackage{epsfig}
\usepackage{bbold}
\usepackage{color}
\usepackage[caption=false]{subfig}

\journal{J. Mech. Phys. Solids}

\newcommand{\be}{\begin{equation}}
\newcommand{\ee}{\end{equation}}
\newcommand{\bea}{\begin{eqnarray}}
\newcommand{\eea}{\end{eqnarray}}
\newcommand{\hz}{\hat z}

\newcommand{\dd}{\partial}

\newcommand{\rd}{{\rm d}}

\newcommand{\kt}{k_{\rm B}T}
\newcommand{\ksp}{k_{\rm sp}}

\newcommand{\sla}{{\sf \Lambda}}

\newcommand{\kBT}{{k_{\textrm{B}} T}}

\newcommand{\LL}{\ell}

\newcommand{\cG}{\mathcal{G}}

\newcommand{\cS}{\mathcal{S}}
\newcommand{\cH}{\mathcal{H}}
\newcommand{\cZ}{\mathcal{Z}}
\newcommand{\cF}{\mathcal{F}}

\newcommand{\Nseg}{N_i}
\newcommand{\Lseg}{\ell_i}
\newcommand{\Ltot}{\ell_{\rm tot}}

\newcommand{\Nab}{N_{\rm AB}}
\newcommand{\InvLgv}{\mathcal{L}_1^{-1}}
\newcommand{\Lgv}{\mathcal{L}_1}

\begin{document}

\begin{frontmatter}




\title{Disorder, pre-stress and non-affinity in polymer 8-chain models}

\author{Adrian R. Cioroianu}
\author{Ewa M. Spiesz}
\author{Cornelis Storm\corref{cor1}}
\ead{c.storm@tue.nl}
\address{Department of Physics and Institute for Complex Molecular Systems, 
Eindhoven University of Technology, 
P.O. Box 513, 5600 MB Eindhoven, The Netherlands}

\begin{abstract}
To assess the role of single-chain elasticity, non-affine strain fields and pre-stressed reference states we present and discuss the results of numerical and analytical analyses of modified 8-chain Arruda-Boyce model for cross-linked polymer networks.
This class of models has proved highly successful in modeling the finite-strain response of flexible rubbers. We extend it to include the effects of spatial disorder and the associated non-affinity, and use it to assess the validity of replacing the constituent chain's nonlinear elastic response with equivalent linear, Hookean springs. 
Surprisingly, we find that even in the regime of linear response, the full polymer model gives very different results from its linearized counterpart, even though none of the chains are stretched beyond their linear regime. 
We demonstrate that this effect is due to the fact that the polymer models are under considerable pre-stress in their ground state. We show that pre-stress strongly suppresses non-affinity in these unit cell models, resulting in a marked stiffening of the bulk response. The effects of pre-stress we discuss may explain why fully affine mechanical models, in many cases, predict the bulk mechanical response of disordered stiff polymer networks so well.

\end{abstract}


\end{frontmatter}

\newpage

\section{Introduction and summary} \label{sec1}
The mechanical response of soft polymeric materials such as rubbers, plastics, gels and filamentous biomaterials is crucial to their technological and biological functioning. While it is obvious that the macroscopic response of these materials is encoded in the behavior of its constituent chains and the manner in which these are connected and arranged in space, {\em predicting} the mechanical response based on such microstructural parameters remains a challenge. Generally, the filaments themselves display strongly nonlinear response to deformations, the spatial arrangement is highly disordered, the strain fields and distributions may be highly heterogeneous and finally, predictive models far beyond linear response are required to capture typical operational conditions for biological and synthetic materials alike. 

Literature abounds with different approaches put forward to tackle one, or several, of these challenges: at the microscopic end of the spectrum, molecular dynamics and Monte Carlo models allow access to micromechanical response using accurate representations of chemical or supramolecular polymer material constituents, but are restricted both in the largest sizes accessible, as well as in the timescales. At the largest scales, phenomenological constitutive models are often able to capture nonlinear bulk response well for specific materials, but lack a direct link to microscopic properties and architecture and therefore must, at some point, invoke effective parameters determined from fits to data. In between the length scales, so-called rubber elasticity models \cite{Flory1989, Rubinstein2007, LangeGent} coarse grain the molecular response into effective ideal \cite{ArrudaBoyce1993,WuGiessen1993, Miehe2004}, wormlike \cite{PalmerBoyce} or differently elastic chain response and, in order to connect to macroscopic response, apply averaging or homogenization techniques to compute the macroscopic response. Each of these approaches has proven extremely useful, and each continues to be improved. From a practical point of view, provided one selects the appropriate approach for a given system and set of mechanical questions, the available spectrum of techniques is eminently capable of accurate, even predictive, analysis. From a fundamental point of view, however, the situation remains unsatisfactory: there is to date no single model that allows predictive modeling of macroscopic mechanical response using as input the single-chain response and the microstructural architecture of the material. Nor is it clear, in fact, what processes and mechanisms such a model should include. Nonlinearities enter at all levels: from the polymers themselves, from spatial and elastic heterogeneity, from composition and geometry. With this paper, we assess the impact of several of these factors - specifically, the elastic nonlinearity of individual polymers, non-affine strain fields, and  pre-stresses - by introducing them, in controlled fashion, into an established framework.  

Our point of departure is a particular class of models: 8-chain or Arruda-Boyce (AB) models. We are aware that numerous improved and more intricate versions have been proposed \cite{ArrudaBoyce1993,WuGiessen1993,KroonNA2,Bergstorm1998,BergstrormBoyce,Mlyniec2014,Cioroianu2013}, but all share the ability to determine the constitutive relation directly from single-polymer response, for a specific microstructural arrangement of chains. 
The AB model shares two important features with classical (Flory) rubber elasticity (CRE) \cite{Flory1989}: It is (macroscopically) isotropic and directly rooted in the response of single, freely jointed chains (FJCs). One important advantage of the AB model is that it explicitly considers the spatial arrangement of these chains, rather than apply to them a statistical averaging as CRE does (its response is computed by averaging over all chain orientations, weighted according to the FJC's Gaussian radial distribution function). In contrast to the CRE model, the AB is a unit-cell model: It comprises eight identical freely jointed chains arranged in symmetric fashion within a cube of linear dimension $a_0$. This basic motif readily permits the introduction of spatial disorder, and numerous more intricate, improved variants of the model exist \cite{ArrudaBoyce1993,WuGiessen1993,KroonNA2}. 
In the following, we study the AB model for different types of polymer elastic behaviour, with the original - ordered - geometry (chains linked in the center of the unit cell) and with positional disorder (chains linked off-center). Our work shows, that there is a direct connection between single-chain elasticity, disorder, non-affinity and pre-stress that has important consequences for the predicted moduli and the response.

Our paper is organized as follows: in Sec. \ref{sec2} we review the Arruda-Boyce model. In Sec. \ref{sec3} we present and  discuss our disordered, non-affine extension of it. 
We demonstrate that the modified model is capable of capturing experimental data. In Sec. \ref{sec4} we compare the stress-strain response in various settings to reveal the importance of pre-stressed states. In Sec. \ref{sec5} we present a measure of non-affinity and compare our findings in both the pre-stressed and stress free ensembles. In Sec. \ref{sec6} we present our conclusions and add some remarks regarding the relevance of pre-stresses in a more general setting and how they affect the degree of affinity of deformations.

\section{Freely Jointed Chain mechanics and the Arruda-Boyce model} \label{sec2}
In the standard Arruda-Boyce 8-chain model \cite{ArrudaBoyce1993} a polymer network is represented as a collection of identical unit cells. Each of these cubic unit cells has a linear dimension of $a_0$ in the undeformed state, and contains 8 polymer chains that connect the cube's vertices to its center. Thus, the 8 chains are identical in initial length $\ell_0$:
\be
\ell_0=\frac12 \sqrt3\, a_0.
\ee
Each of these chains is modeled as a Freely Jointed Chain (FJC): the well-known model for flexible (Gaussian) polymers \citep{Rubinstein2007,Flory1989,Boal}. The FJC represents a long polymer with a contour length $\Ltot$ as a sequence of $N$ connected, but freely hinged, Kuhn segments each of which has a length $b$ (such that $N b = \Ltot$), where $b$ is chosen several times larger than the persistence length of the polymer to ensure that no orientational correlations between the segments remain.

The FJC has no internal energy - each of its conformations may be transformed into any of the other conformations at no energetic cost. Its Gibbs free energy, in the presence of an external applied force $f$ in the $\hz$-direction, is therefore
\be
\cG=\cH-T\cS=-f \LL_z -T\cS\, ,
\ee
where $\cH$ is the enthalpy, $T$ is the absolute temperature, $\cS$ is the entropy and $\LL_z$ is the end-to-end length of the polymer in the direction of the force, {\em i.e.} it is the $\hz$ component of the full end-to-end vector $\vec \LL$ which - expressed in terms of the unit bond vectors $\hat t_i$ - is computed as
\be
\vec \LL=b\sum_{i=1}^{N}\hat t_i\, .
\ee
The partition function in the Gibbs ensemble may be computed as 
\be
\cZ(N,f,T)=\int\!\! \rd \hat t_1 \cdots \int\!\! \rd \hat t_N \, \exp\left(\frac{f b}{\kt}\sum_{i=1}^{N}\hat t_i\cdot \hz\right)
\ee
In three dimensions, the partition function is readily computed and from it, an analytical expression for the Gibbs free energy is obtained
\bea
\cG(N,f,T)&=&-
\kt \log \cZ(N,f,T)\nonumber \\&=&-N \kt \log\left(\frac{4 \pi \sinh(fb/\kt)}{fb/\kt}\right)\, .
\eea
We may perform a Legendre transformation on $\cG(N,f,T)$ to obtain from this the Helmholtz free energy $\cF=\cG+f \LL_z$ to find for the free energy of a FJC subject to a force $f$
\be\label{hfe}
\cF(N,\LL_z,T)=-N \kt \log\left(\frac{4 \pi \sinh(fb/\kt)}{fb/\kt}\right)+f \LL_z\, .
\ee
Evaluating now the condition that $\cF$ be minimal in equilibrium yields the mechanical response of the FJC, summarized as the relationship between its projected end-to-end length $\LL_z$ and the externally applied force $f$ \cite{Rubinstein2007} 
\bea\label{fe1}
\frac{\dd \cF}{\dd f}&=&0 \Rightarrow \nonumber \\
\langle \LL_z\rangle(f)&=&N b \left[\coth\left(\frac{f b}{\kt}\right)-\frac{\kt}{fb}\right]\nonumber \\
&\equiv& N b \Lgv\left(\frac{f b}{\kt}\right)\, ,
\eea
where we define the first order Langevin function: $\Lgv(x)=\coth x-x^{-1}$. The angular brackets, $\langle \cdot \rangle$, signify ensemble averages. In the following, we will denote by $\LL$ this ensemble average and will therefore drop the brackets as well as the subscript $z$. While the Langevin function does not have an analytical inverse, we may use Eq. (\ref{fe1}) to formally extract the force-extension relationship that we will use throughout this paper
\be\label{fe}
f(\LL)=\frac{\kt}{b}\InvLgv\left(\frac{\LL}{Nb}\right)\, .
\ee
Since we shall generally specify the state of {\em strain}, rather than the stress, it is convenient to substitute the expression for $f(\ell)$ in to the formula for the Helmholtz free energy Eq. (\ref{hfe}), to arrive at the expression also used in  \cite{ArrudaBoyce1993}
\bea\label{ener}
\!\!\!\cF(N,\ell,T)\!\!&=&\!\!N \kt\left[\left(\frac{\ell}{N b}\right)\!\!\InvLgv\!\!\left(\frac{\ell}{Nb}\right)+\right. \nonumber \\ && \left. +\!\log\!\left(\frac{\InvLgv\!\!\left(\frac{\ell}{Nb}\right)}{\sinh\left(\InvLgv\!\!\left(\frac{\ell}{Nb}\right)\right)}\right)\right]\!\!+\!\!\cF_0(N,T)
\eea
We may expand this expression for small $\ell$, using the Taylor expansion of the inverse Langevin function $\InvLgv(x)\approx 3 x$, to find that
\be\label{enerH}
\cF(N,\ell,T)\approx \frac12 \left(\frac{3 \kt}{N b^2} \right) \ell^2\,
\ee
From which we see that the elasticity, for small extensions, is Hookean with an entropic spring constant equal to 
\be\label{kspH}
k_{\rm sp}=\frac{3 \kt}{N b^2}. 
\ee
The equilibrium length is zero for the FJC. 

We now turn to the description of deformation applied to the unit cell. We will start with a deformation that comprises three principal stretches $\lambda_i$, and transforms the boundary according to
\be
\sla =\left(
\begin{array}{ccc}
 {\lambda_{1}} &  0 & 0 \\
 0 &  {\lambda_{2}} & 0 \\
 0 &  0& {\lambda_{3}}
\end{array}
\right)\, ,
\ee
which maps each point $\vec r_b$ on the boundary of the unit cell to $\vec r'_b=\sla\cdot\vec r_b$. Since the arrangement of chains within the Arruda-Boyce unit cell is fully symmetric, we may infer from the deformation of the boundary also the displacement of the central point, where the eight chains meet. This is a crucial point: by construction, the Arruda-Boyce unit cell is constrained to deform {\em affinely}.

This considerable simplification means that specifying $\{\lambda_1,\lambda_2,\lambda_3\}$ gives full knowledge of the deformed lengths of all eight chains. 
Moreover, because of their symmetric arrangement in space, in this deformation all chains experience identical changes in length, and the new (deformed) projected end-to-end length and the original length is given by
\be
\ell(\lambda_1,\lambda_2,\lambda_3)=|\sla\cdot\ell_0|=\frac{a_0}{2}\left(\lambda_1^2+\lambda_2^2+\lambda_3^2\right)^{1/2}\, ,
\ee
where $a_0$ is the linear dimension of the unit cell and $\ell_0$ is the initial length of the chain. In the AB model, the eight chains are modeled as FJC's and the undeformed lengths are identified with the root of the mean-squared end-to-end length
\be
\ell_0=\sqrt{\langle \ell^2(f=0) \rangle}=\sqrt{N} b
\ee
This identification has important consequences: the undeformed length is {\em not} the length at zero force, which is zero as is also evident from Eq. \ref{fe1}. Thus, the chains in the AB model need to be stretched in order to reach the central point, by a tension which may be computed to be
\be
f_{PS}=\frac{\kt}{b}\InvLgv\left(\frac{1}{\sqrt{N}}\right)\, .
\ee
This gives rise to a pre-stress in the resulting model which we will return to in the following. With this definition, however, the length of the chains may be expressed in terms of the FJC parameters $N$ and $b$, and the principal stretches $\lambda_i$, as
\be\label{forsub}
\ell(\lambda_1,\lambda_2,\lambda_3)=\frac{1}{\sqrt{3}}\sqrt{N} b \left(\lambda_1^2+\lambda_2^2+\lambda_3^2\right)^{1/2}\, .
\ee
By substituting Eq. \ref{forsub} into Eq. \ref{ener}, we now have an expression for the free energy density of deformation expressed in terms of the stretches $\lambda_i$ and the FJC parameters.
\be
f(\lambda_i)=\rho \cF(N,\frac{1}{\sqrt{3}}\sqrt{N} b \left(\lambda_1^2+\lambda_2^2+\lambda_3^2\right)^{1/2},T).
\ee
where $\rho=8/a_0^3$ is the chain density for the eight-chain AB model in the cube of side $a_0$. From this, we may compute the principal (Cauchy) stress tensor components
\be
\sigma^{C}_i=\frac{1}{\det \sla}\left(\frac{\dd f}{\dd \sla_{ii}}\right)(\sla^T)_{ii}= \left(\frac{\lambda_i}{\lambda_1 \lambda_2 \lambda_3}\right)\frac{\dd f}{\dd \lambda_i}+p\, ,
\ee
Where $p$ is an arbitrary hydrostatic pressure, which may be applied to ensure the pre-stresses are compensated for in the reference configuration. For incompressible materials, the determinant of $\sla$ equals one and we find, by direct differentiation, and after some algebra
\bea\label{sigc}
\sigma_i^C&=&\frac{1}{3}\left(\rho \kt\right)\sqrt{N}\left(\frac{\sqrt{3}\lambda_i^2}{\left(\lambda_1^2+\lambda_2^2+\lambda_3^2\right)^{1/2}}\right)\times\nonumber \\
& & \times \, \InvLgv\left(\frac{\left(\lambda_1^2+\lambda_2^2+\lambda_3^2\right)^{1/2}}{\sqrt{3N}}\right)-p\,.
\eea
The combination $\rho \kt$ is commonly denoted $C_R$, the classical rubber elastic modulus (although we note that in their original paper, Arruda and Boyce denote by CR the full prefactor $\frac13 \rho \kt$).
In this paper, we will follow conventions and report {\em nominal}, rather than Cauchy stresses - these are the forces per unit underformed area, which we shall denote by simply $\sigma$. 
For an incompressible material they may be straightforwardly obtained from $\sigma^C$ by correcting for the incurred stretches as
\be
\sigma_i=\left(\frac{1}{\lambda_i}\right)\sigma^C.
\ee

\section{Model definitions and simulation protocol} \label{sec3}

Throughout this paper we will analyze and compare nominal stresses for various types of homogeneous deformations:
\begin{itemize}
 \item uniaxial extension or compression, for which the deformation gradient tensor is
\be\label{eqUA}
\sla_U=\left(
\begin{array}{ccc}
 \lambda  & 0 & 0 \\
 0 & \lambda^{-\frac{1}{2}}  & 0 \\
 0 & 0 & \lambda ^{-\frac{1}{2}} \\
\end{array}
\right)
\ee
 \item biaxial extension or compression, for which
\be\label{eqBA}
\sla_B=\left(
\begin{array}{ccc}
 \lambda^{-2}  & 0 & 0 \\
 0 & \lambda  & 0 \\
 0 & 0 & \lambda \\
\end{array}
\right)
\ee
 \item simple shear, for which
\be
\sla_S=\left(
\begin{array}{ccc}
 1  & 0 & \lambda \\
 0 & 1  & 0 \\
 0 & 0 & 1 \\
\end{array}
\right)
\ee
 \item  pure shear (or plane strain compression) for which
\be
\sla_P=\left(
\begin{array}{ccc}
 \lambda  & 0 & 0 \\
 0 & 1  & 0 \\
 0 & 0 & \frac{1}{\lambda}  \\
\end{array}
\right)
\ee
 \item  uniform expansion or compression. This deformation mode does not conserve volume and we will use to assess the hydrostatic pre-stress which is probed by changing the volume.
\be
\sla_V=\left(
\begin{array}{ccc}
 \lambda  & 0 & 0 \\
 0 & \lambda  & 0 \\
 0 & 0 & \lambda  \\
\end{array}
\right)
\ee
\end{itemize}
We point out that Eqs. \ref{eqBA}-\ref{eqUA} are in fact inter-changeable provided that $\lambda^{-2}\rightarrow\lambda$ or $\lambda^{-\frac{1}{2}}\rightarrow\lambda$, but for the sake of consistency with previous work we will keep both definitions from now on.
Our first objective is to use the AB modeling approach to assess the dependence of the predicted mechanical response on the type of elasticity chosen for the individual polymers: Hookean springs with spring constants as defined in Eq. \ref{kspH}, and Freely Jointed Chains with energy is given by Eq. \ref{ener}. 
For the linear elastic Hookean springs, as is commonly done in polymer network modeling \cite{Head2003a, Head2003c, Gardel2004, Conti2009}, we assume a rest length equal to the underformed length.
The energy of linear elastic spring is equal to $\frac{1}{2} \ksp \delta\ell^2$. Taking the linearized spring constant of freely jointed chain model for finite deformations (Eq. \ref{kspH}) we arrive at:
\be\label{enerH2}
\cF(N,\ell,T)\approx \frac12 \left(\frac{3 \kt}{N b^2} \right) \delta\ell^2\, ,
\ee
with $\delta \ell$ now the extension away from the rest length $\ell_0$. 
In the following we test whether replacing the nonlinear force-extension of the FJC model with its linear, finite rest-length approximation affects the linear (and nonlinear) response of an AB-type network.

\begin{figure}
\includegraphics[width=.45\linewidth]{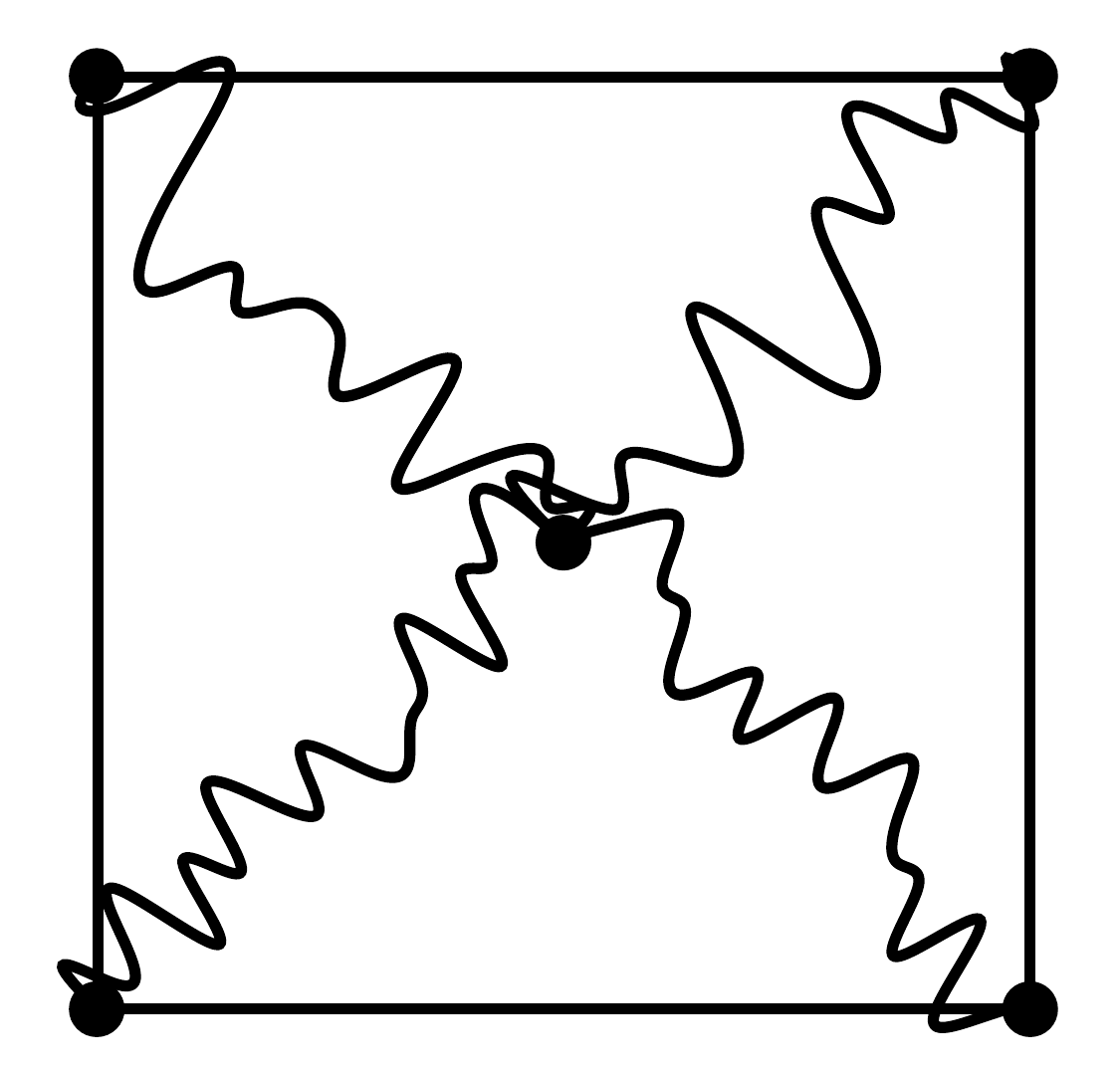}
\includegraphics[width=.45\linewidth]{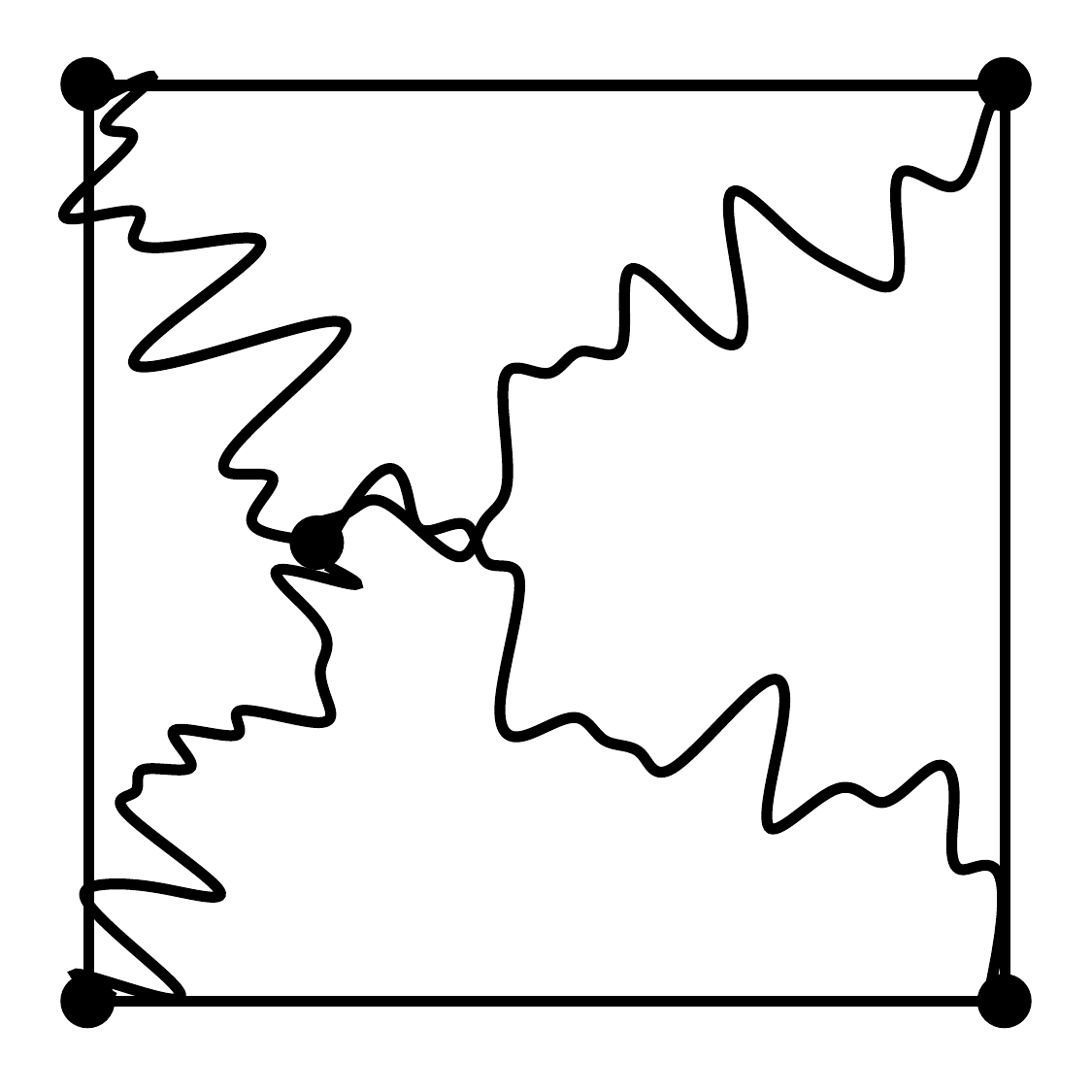}
\caption{\label{sqoffC}Schematic two dimensional representation of a typical Arruda-Boyce setting (left) and our model incorporating positional disorder with the off-center linker (right).}
\end{figure}

The second objective is to study the effects of allowing for non-affine deformations. 
In order to do this, we modify the geometry, placing the linker connecting the eight chains off-center (this is schematically shown in a 2D plot in Fig. \ref{sqoffC}).

This introduces important degrees of freedom into the AB model: the asymmetric (off-center) placement of the linker allows non-affine deformations of the connected chains. 
That is, if its initial position is $\vec r_c$, its position after the boundaries of the cube have been subjected to $\sla$ is {\em not} $\sla\cdot\vec r_c$. 
Rather, the new (deformed) position is determined by minimizing the sum of the elastic energies from all eight chains. The elastic energies thus obtained are then averaged among all the disordered unit cells considered.
This introduces elastic inhomogeneity within the unit cell. When the linker is not centrally placed, some chains are shorter than others and therefore stiffer than others: 
As may be surmised from Eq. \ref{enerH}, the spring constant for entropic FJC's scales as $N^{-1}$, and longer chains are thus less stiff (lower spring constants imply less force is required for the same extension) than shorter ones. 
We shall term the first effect {\em positional disorder}, and the second {\em elastic disorder}. One final effect that is unavoidably incurred is related to the density. 
Eight chains connected with a linker positioned exactly in the center of the unit cell requires the least amount of spring contour length, and thus configurations that are linked elsewhere inside the volume require more spring length per unit volume.

In what follows, we focus on elastic and positional disorder. For the polymer models we are considering, the two are inseparable, as the length differences that give rise to positional disorder are directly related to stiffness differences. 
In order to disentangle these from density effects, we have devised a computational method that allows one to work in a constant density ensemble. For FJCs this has important consequences for mechanical response as we will show. 

Previous work has examined the effects of either non-affinity of the boundaries (modeling the compliance of the surroundings of the quasi-unit cell) \cite{KroonNA1}, frozen positional disorder (the case where the interior linker point is distributed throughout the unit cell, but still constrained to deform affinely) \cite{WuGiessen1993}, and coarse-grained (homogenized) approaches to the average non-affine displacements \cite{Miehe2004}. 
These works emphasize the large effects of both types of disorder on mechanical response, but each is unable to disentangle these effects from either the constitutive behavior of the chains themselves, or from density effects. This is the central novelty of our work: by operating in a constant density ensemble, averaging over well-defined discrete networks, we are able to clearly establish the contributions of polymer response and non-affinity.

The constant density ensemble is constructed as follows. We start from the observation that each AB unit cell, as defined in the original paper, contains 8$\times N$ FJC segments, each of length $b$. 
In all of our simulations, we set $a_0$ to a value of 1 (we will work in non-dimensional units, scaling lengths by $a_0$ from now on, but note that when fitting to data the value of CR or, alternatively, $\rho \kt$, sets all relevant length- and force scales) and $N=50$. 
The task at hand is to construct all configurations that contain exactly 400 segments, distributed in all possible manners across the eight chains. 
We do this by first partitioning the 400 segments into eight subsets $\{N_i\}_{i=1}^8$, constructing FJC's out of each. 
We then presume that these eight chains are connected at some position $\vec r_c$ within the cube by solving the force balance equation. $\vec r_c( \{N_i\})$ is thus defined by the implicit equation
\be\label{fB}
\sum_{i=1,8} \vec f_{i}(\vec r_c)=\left(\frac{\kBT}{b}\right)\sum_{i=1,8}\hat \Lseg(\vec r_c)\, \InvLgv \left(\frac{\Lseg(\vec r_c)}{b \Nseg}\right)=\vec 0.
\ee
Here, $\hat \Lseg(\vec r_c)$ is the unit vector pointing along polymer $i$. This force balance relation consists of three separate force balances along the three axes. 
Since there are three degrees of freedom (the coordinates $\vec r_c$), this equation has a unique solution. 

The force balance criterion described by Eq. \ref{fB} can be achieved by following any of the two subsequent routes: firstly, place the linker at a desired position $\vec r_c$ and retrieve the set $N_i$ that satisfies Eq. \ref{fB}; and secondly, choose a set $N_i$, under the constant density condition, and compute the corresponding position $\vec r_c( \{N_i\})$ of the crosslinker. The former route would be the path of choice as it would allow to tune the resulting configuration to any desired spatial distribution. This former route is, unfortunately, mathematically unfeasible, as there are eight unknowns to be retrieved from a system of at most four equations: three equations dictating the force balance criteria for each Cartesian coordinate and a fourth equation for the constant density ensemble.

\begin{figure}
\includegraphics[width=.5\linewidth]{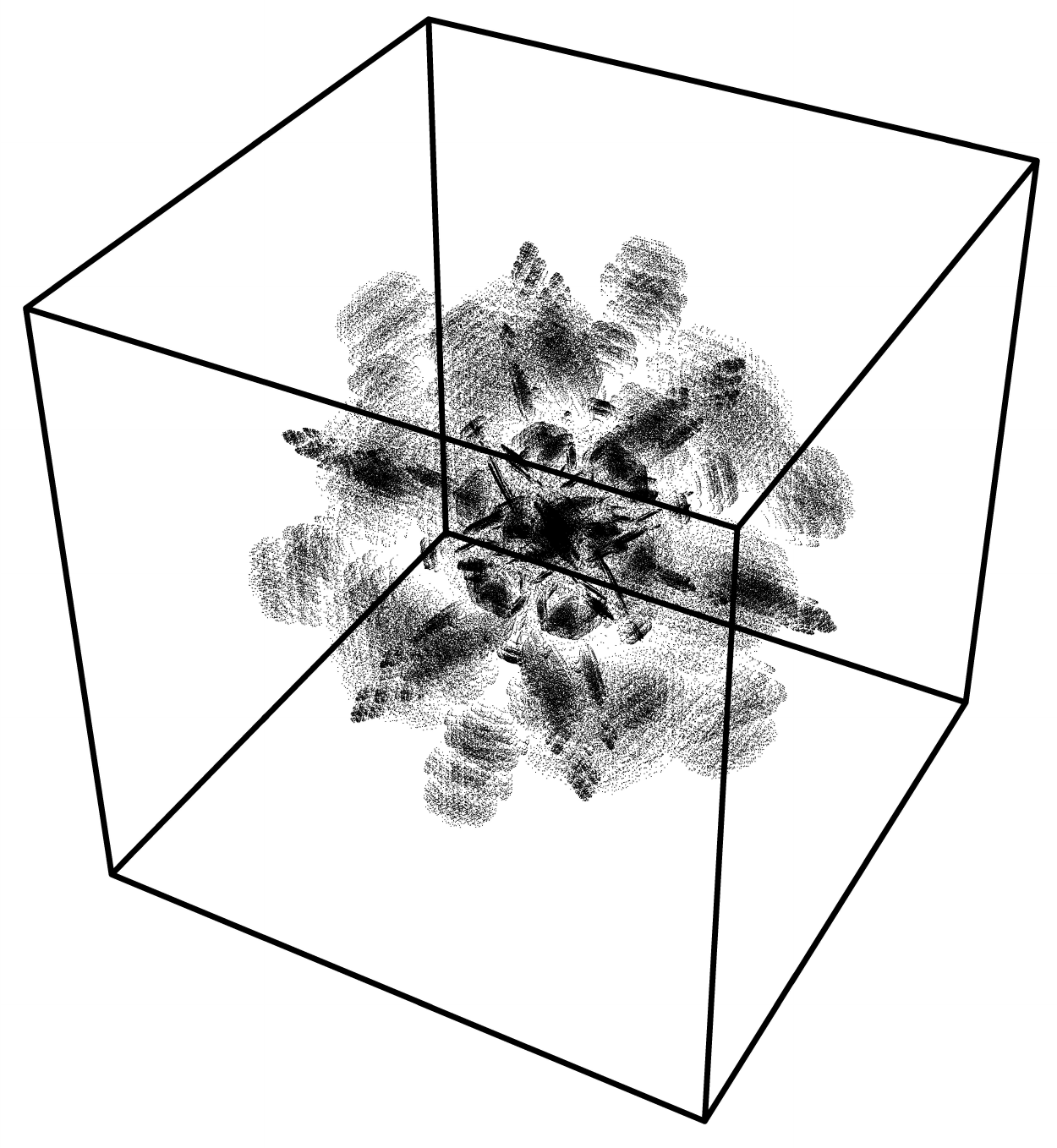}
\includegraphics[width=.5\linewidth]{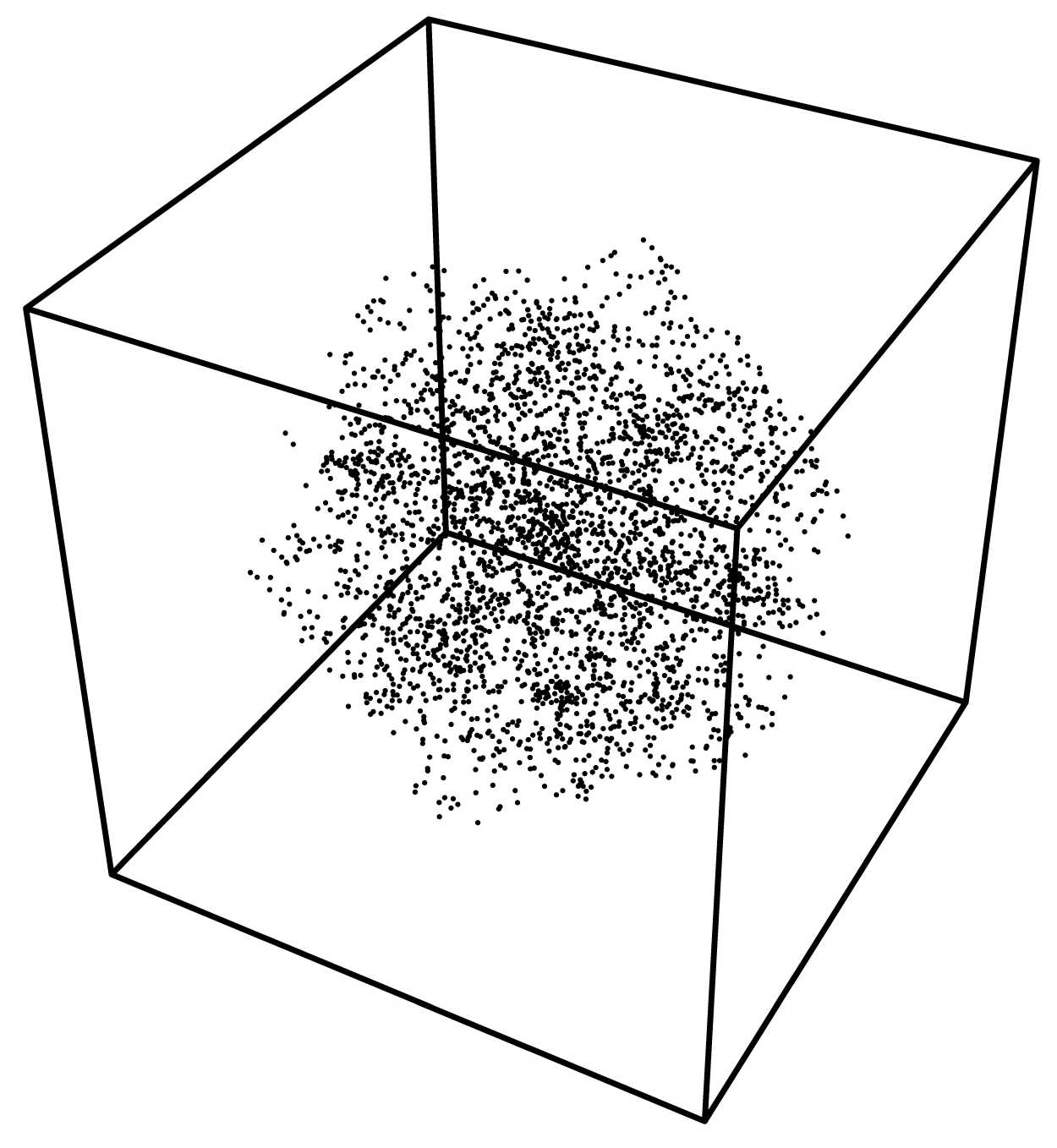}
\caption{\label{cubeDist}The ensemble of equal density networks (left panel) that are generated with the force balance equation is further subjected to a quasi-random selection (right panel) to yield a smooth spatial distribution of the intersection point.}
\end{figure}

We therefore turn to the latter route and start by randomly picking a set $N_i$, solve Eq. \ref{fB}, store the position of the crosslinker $\vec r_c( \{N_i\})$ and repeat this procedure to yield a spatial distribution similar to the one graphed in the upper panel of Fig. \ref{cubeDist}.
Importantly, though all these equilibrated ground state configurations with off-center linkers $\vec r_c( \{N_i\})$ have exactly the same density, the differ greatly in the amount of {\em pre-stress}. In a single chain framework, we define pre-stress as the amount of tension needed to maintain the chain's ends at specific points: one end is undoubtedly tethered at one vertex of the unit cell and the other end is placed at $\vec r_c( \{N_i\})$.
Each possesses internal forces that sum to zero, but as all of the FJCs are stretched away from zero length, the tension in each chain is not equal to zero. 
As we have seen in the previous, the stiffness of chains increases with their length and to avoid physically unrealistic short chains (which contribute disproportionally to the overall response) we place lower-bound cutoff of 10 segments on $N_i$.

The representation of linker positions obtained in the described manner is shown in Fig. \ref{cubeDist}, with the upper panel plotting the full set of $\vec r_c( \{N_i\})$ that meet the force balance criterion and the lower panel with a quasi-random selection. 
As one can see, the full ensemble is highly textured and the density of points is not constant throughout the unit cube. We infer that this textured structure is an immediate result of the nonlinearities present in the force-extension curve of individual chains.
As this texture is a result of our procedure, and does not reflect a physical property of a random sub volume of polymer material, we do not sample randomly from this ensemble but rather do so correcting for its density variations: we subdivide the unit cube into smaller volumes (one thousand cubical boxes, each having an edge length equal to $1/10$ of the unit cell's edge length) and pick randomly ten possible configurations from within each randomly chosen sub volume, to distribute the sampling points evenly across the unit cube. 

Once an equilibrium configuration is chosen, it is subjected to the various deformations detailed above. 
During the deformation, we introduce on last feature to the procedure: we either allow the linker to displace non affinely, or we force it to deform affinely. Thus, we study six distinct variants of Arruda-Boyce type unit cell models, evaluated at identical densities but with or without nonlinear stretching, with or without positional/elastic disorder and with or without the freedom to deform affinely. These models are collected Table 1. We now proceed to analyze the differences and similarities among them.

\begin{table*}[t]
 \caption{The models compared in this paper} \label{deffjc1}
 \begin{tabular}{l l l l l}
 Model & Linker Placement & Affine/Nonaffine & Chain Elasticity & pre-stressed\\
 \hline
Polymer Arruda-Boyce (PAB) & central & affine by symmetry & FJC & yes\\
Hookean Arruda-Boyce (HAB) & central & affine by symmetry & Hookean & no\\
Polymer Non-Affine (PNA) & distributed & non-affine & FJC & yes\\
Hookean Non-Affine (HNA) & distributed & non-affine & Hookean & no\\
Polymer Disordered Affine (PDA) & distributed & forced affine & FJC & yes\\
Hookean Disordered Affine (HDA) & distributed & forced affine & Hookean & no
 \end{tabular}
 \end{table*}
 
\begin{figure}
\includegraphics[width=.9\linewidth]{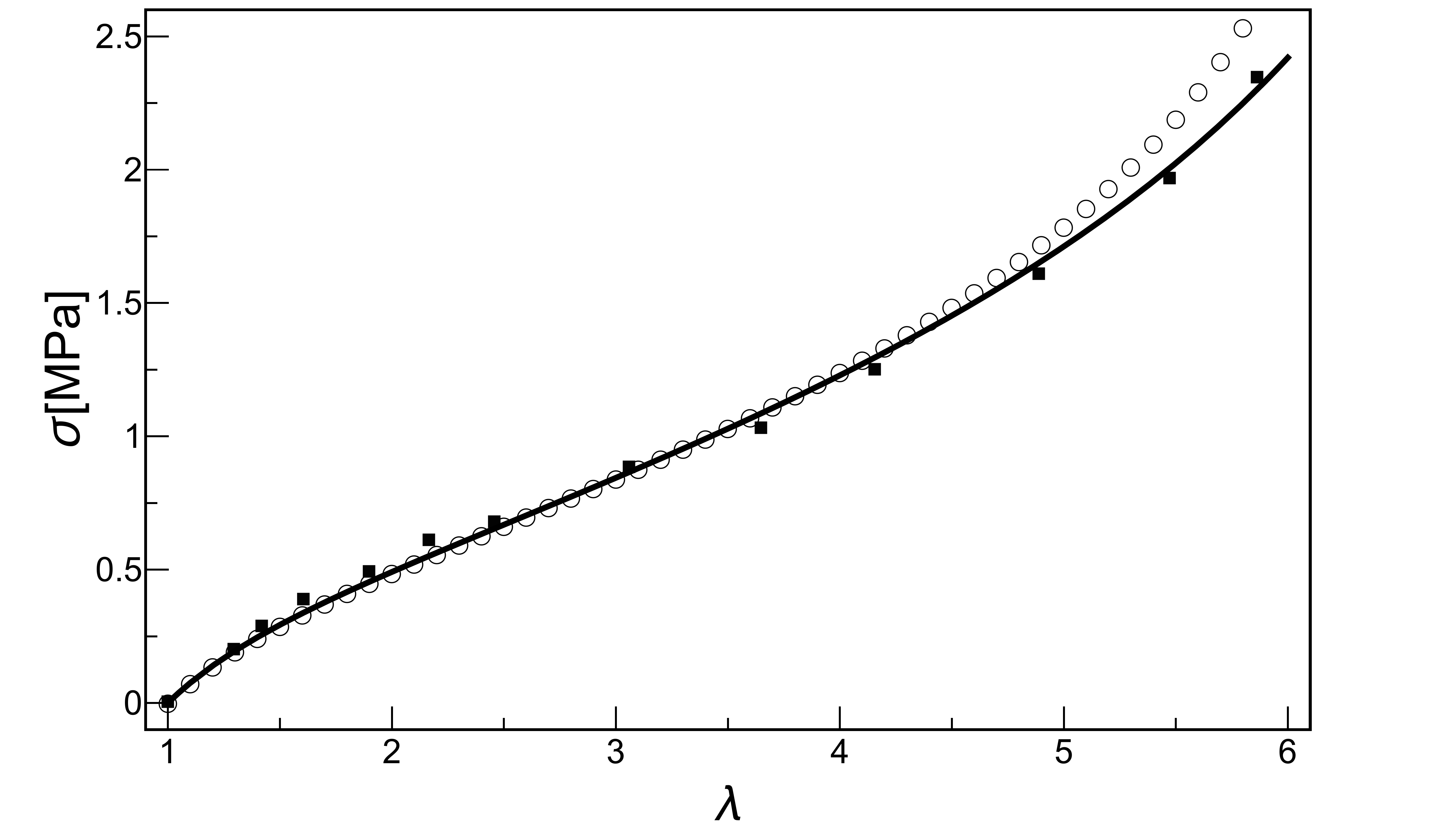}
\caption{\label{fitstress}Stresses $\sigma$[MPa] versus applied strain $\lambda$ in an uni-axial type of deformation. Averaged stresses $(\bigcirc)$ in a network assembled by modified disordered unit cells plotted against experimental data $(\blacksquare)$ obtained in \cite{Treloar1944} and used in the original Arruda-Boyce model \cite{ArrudaBoyce1993}. Both data sets are graphed alongside the analytical curve (black line) predicted by the Arruda-Boyce model \cite{ArrudaBoyce1993}. The presence of shorter chains in the statistics increases stress values for large deformations.}
\end{figure}

As a first check, we compare the ability of the full non-affine unit cell model PNA to that of the original Arruda-Boyce model to capture the data of Treloar. The results are shown in Fig. \ref{fitstress}, which demonstrates that with the appropriate parameter choices, ($a_0=0.5945$, $N_i=212$ and $b=0.1$) our model is equally capable of describing the data, with an equal amount of fit parameters but including the microscopic effects of non-affinity.
We note that the parameters for which this fit was obtained were $CR=0.09$MPa, and $N=26.5$ \cite{ArrudaBoyce1993}. 
While we are generally interested in the generic behavior of modified AB models, we do note that as the AB model is one of the few that allows one, in principle, to relate macroscopic stresses directly to microscopic response these are not simply fit parameters: they do pertain to the actual densities, at least those obtained within the context of the model. 
The fits obtained for the Treloar data, for instance, translate into a unit cube with a side of about 5nm, and a link length $b=0.83$nm. 
This is very close to molecular length scales and it is highly questionable whether FJC's remain credible representations of rubber-like networks at these length scales.

\section{Comparing the stress-strain response of the six models} \label{sec4}

\begin{figure*}
\subfloat[uniaxial deformation - energies]{\includegraphics[width = 7cm]{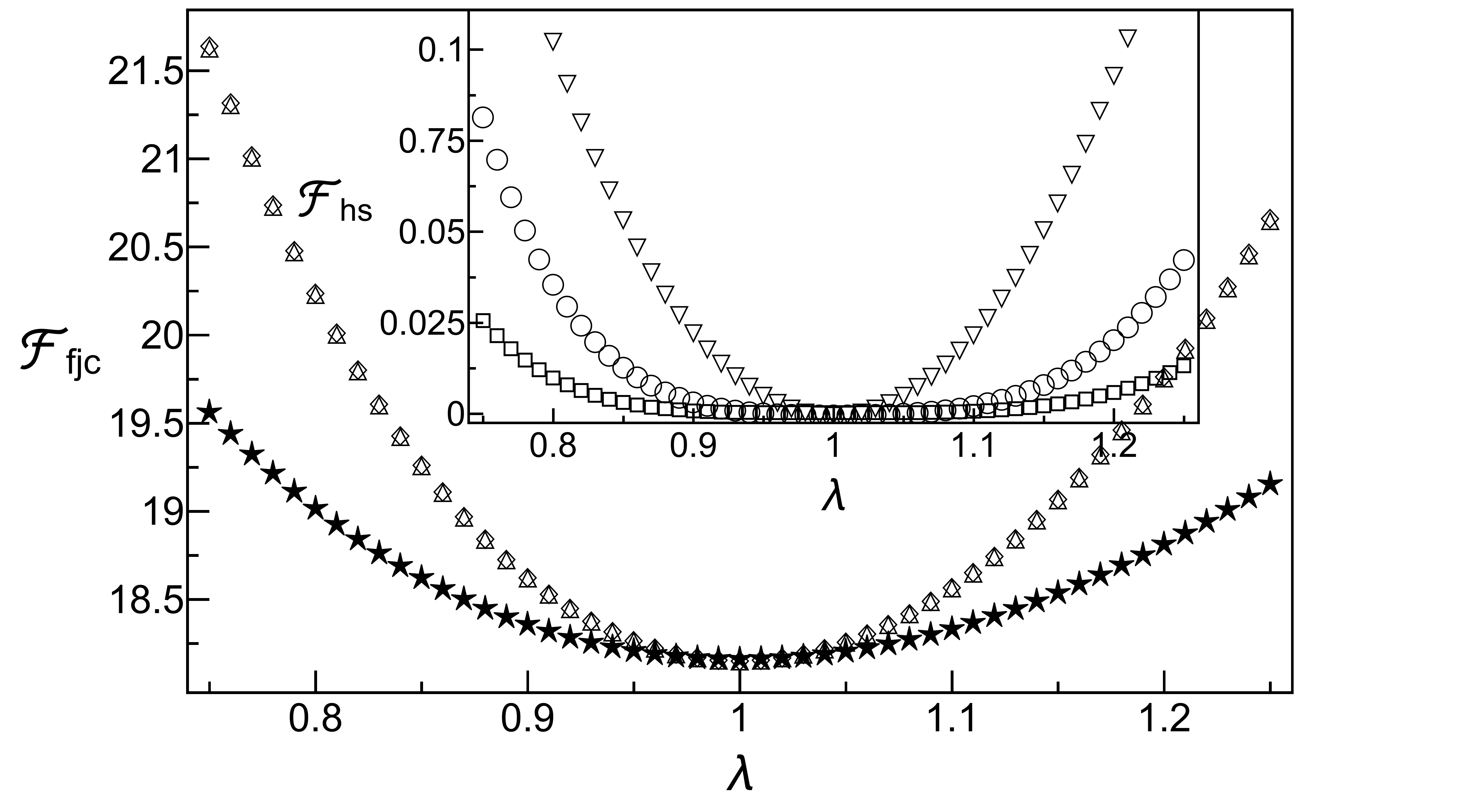}}
\subfloat[uniaxial deformation - stresses]{\includegraphics[width = 7cm]{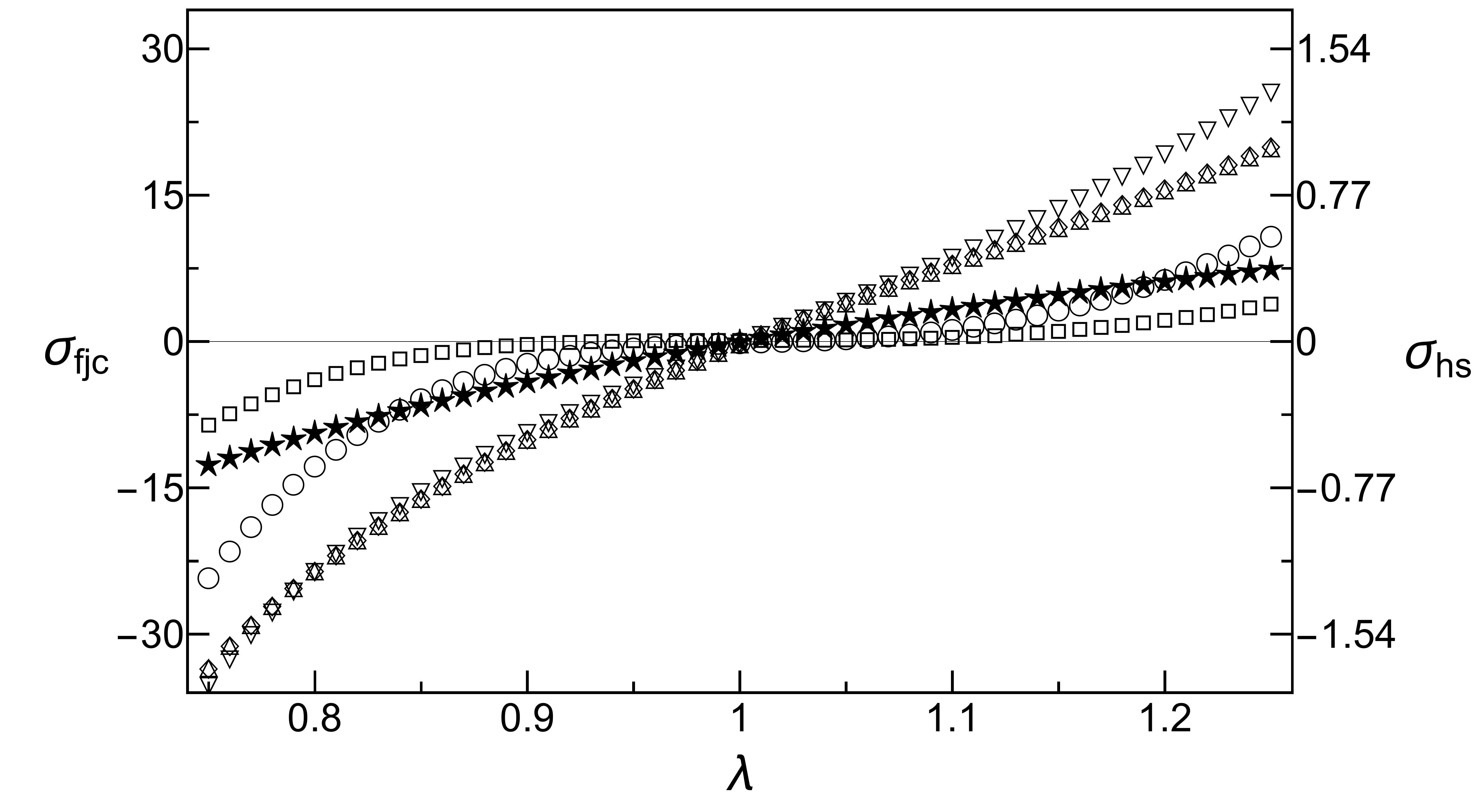}}

\subfloat[biaxial deformation - energies]{\includegraphics[width = 7cm]{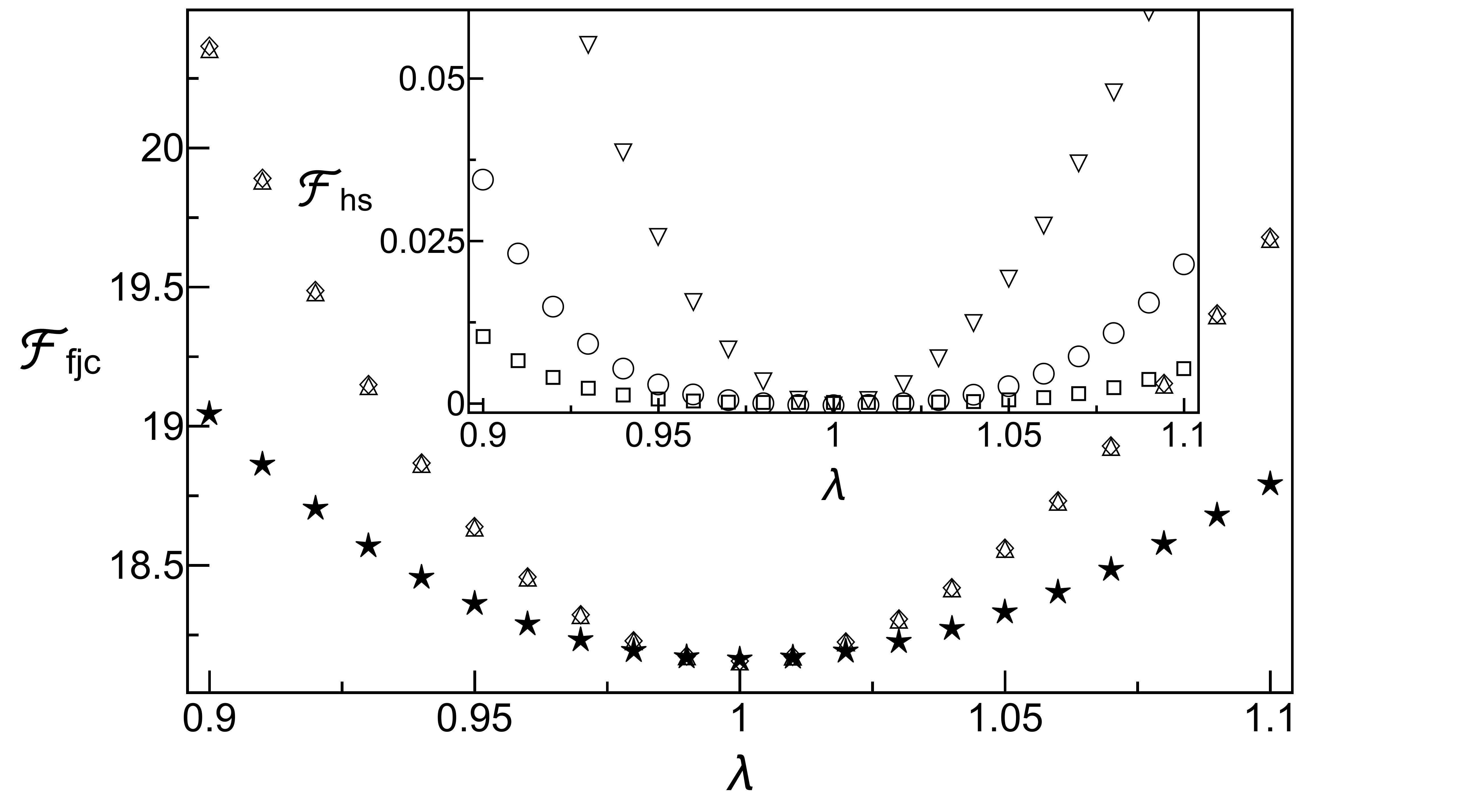}}
\subfloat[biaxial deformation - stresses]{\includegraphics[width = 7cm]{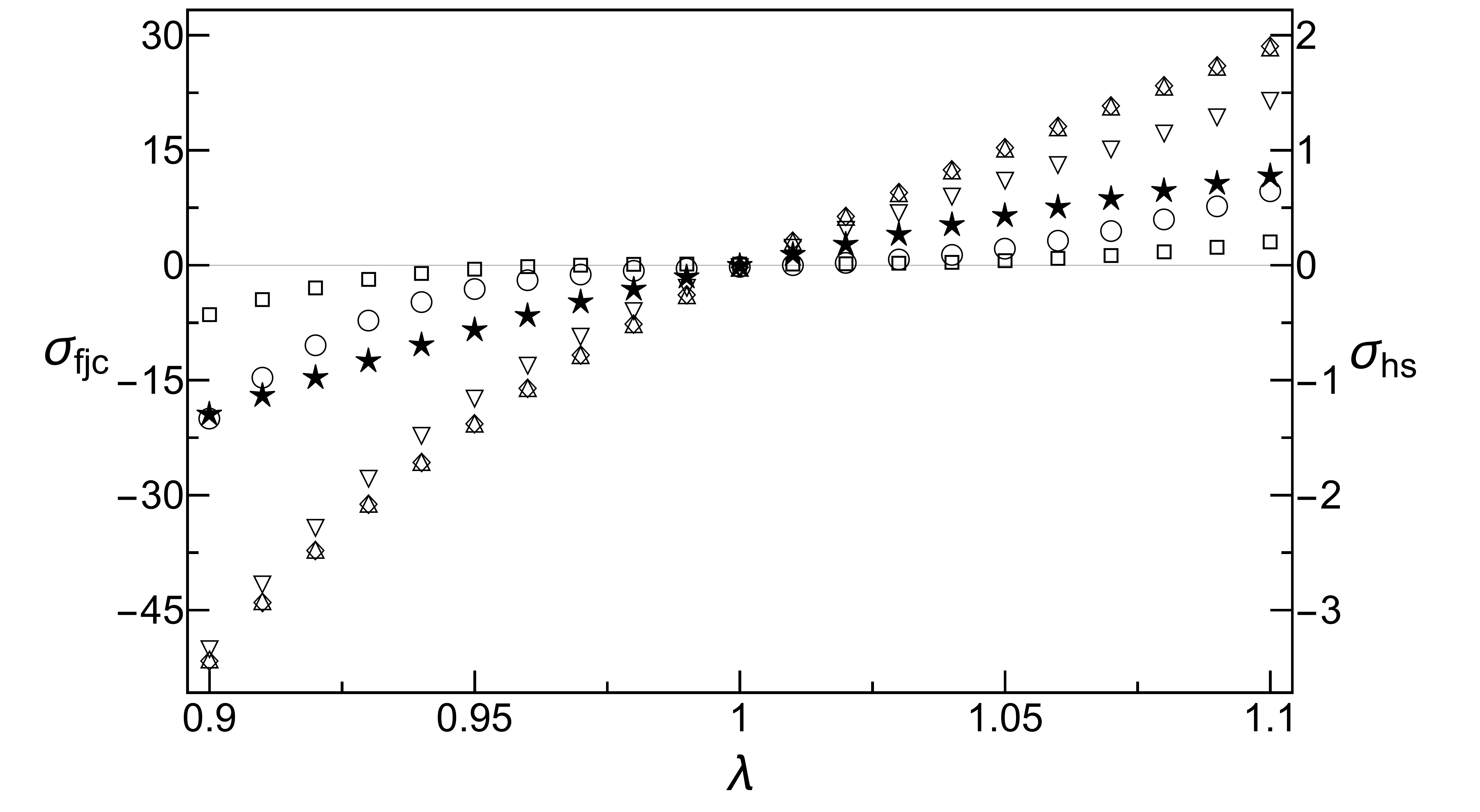}}

\subfloat[simple shear - energies]{\includegraphics[width = 7cm]{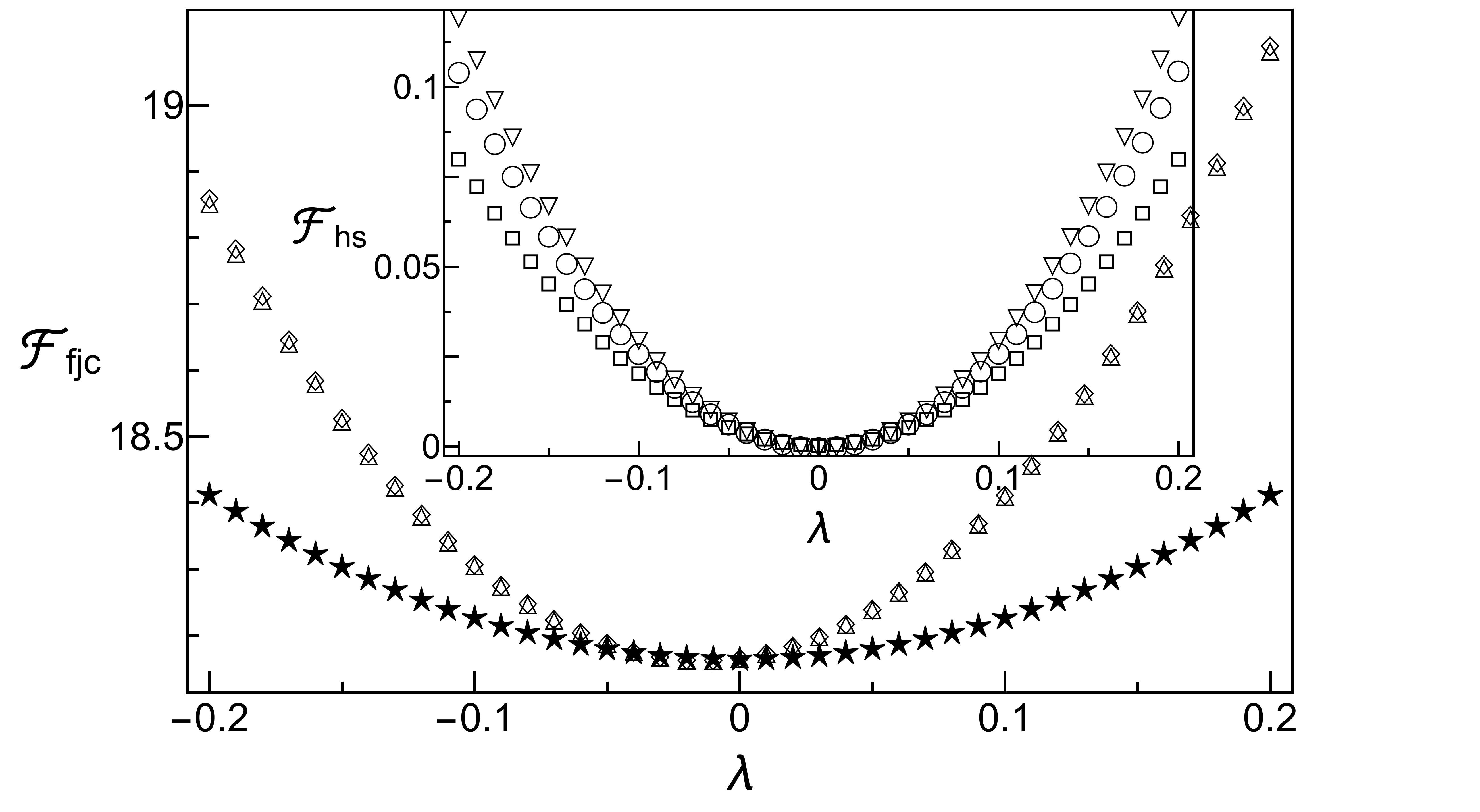}}
\subfloat[simple shear - stresses]{\includegraphics[width = 7cm]{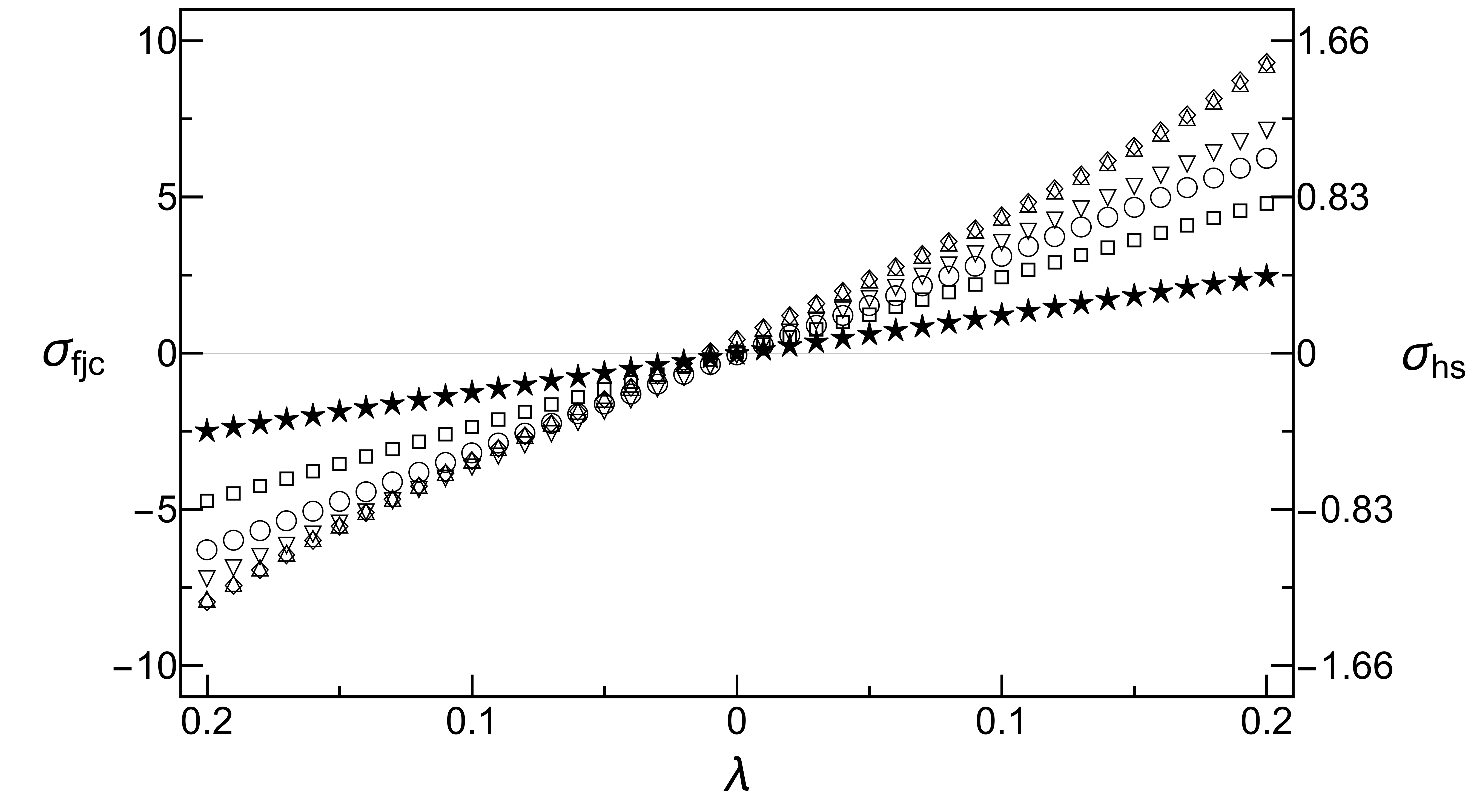}}

\subfloat[pure shear - energies]{\includegraphics[width = 7cm]{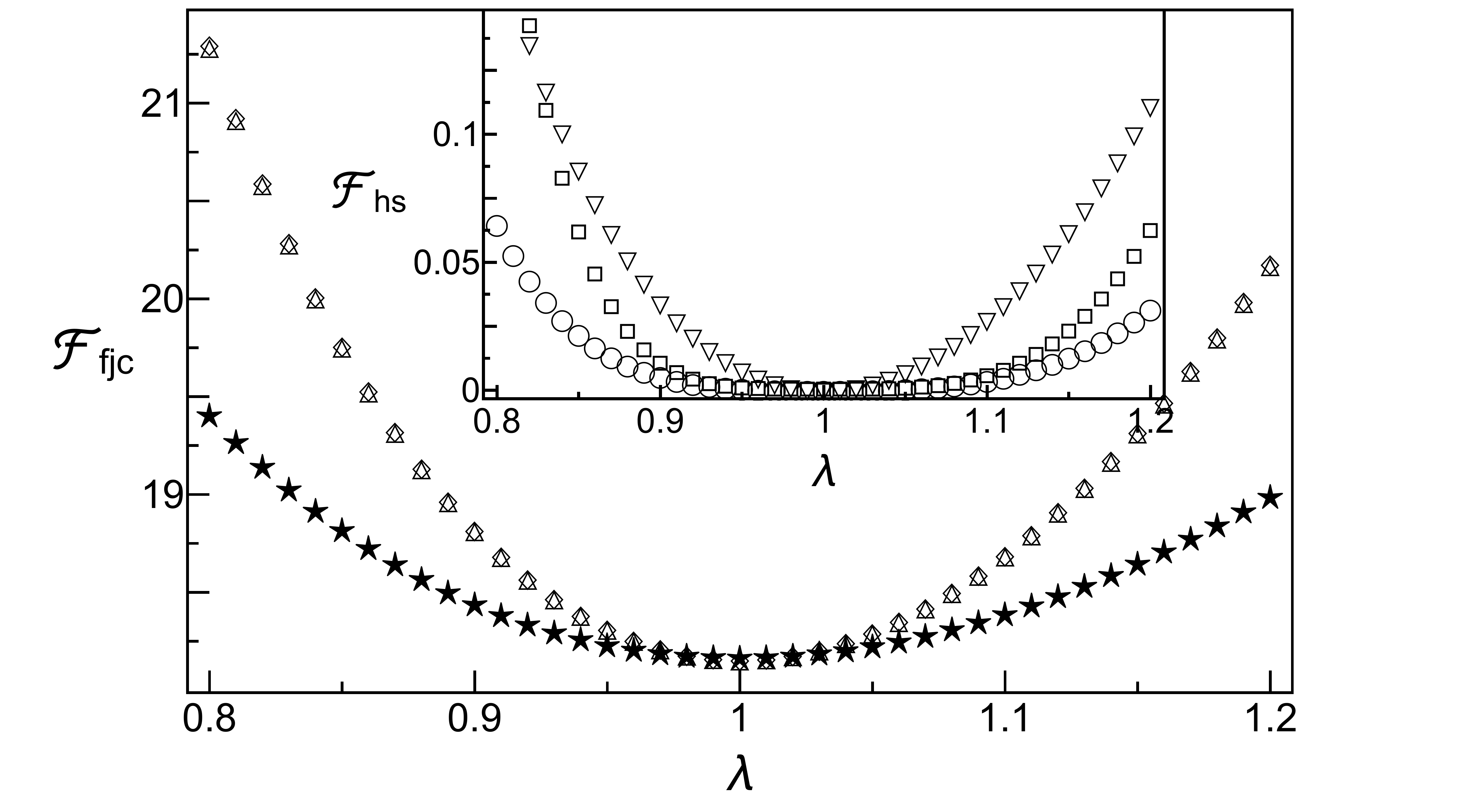}}
\subfloat[pure shear - stresses]{\includegraphics[width = 7cm]{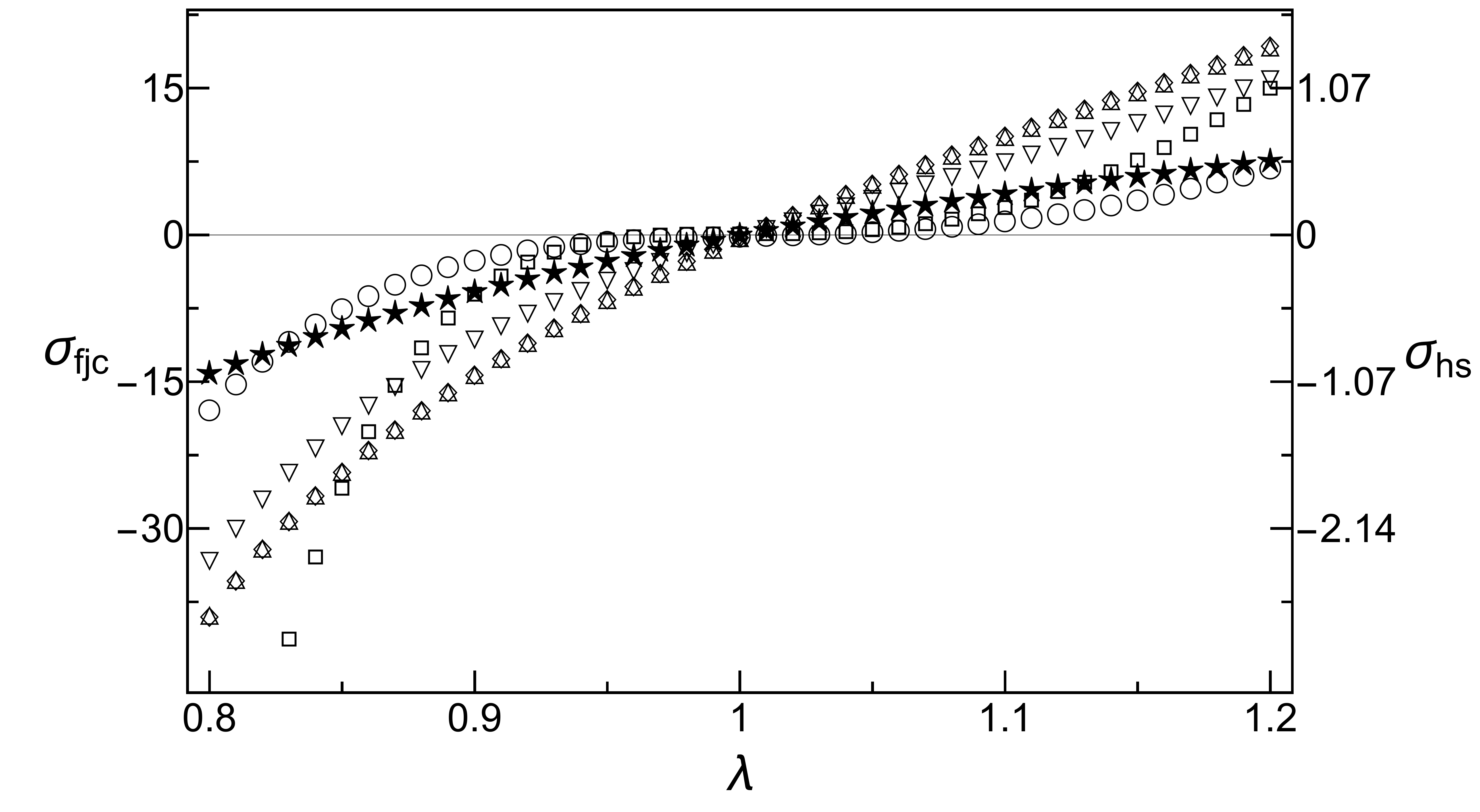}}

\caption{\label{eua}Energies (left column) and stresses (right column) versus applied deformation $\lambda$ in the homogenous deformations used. In the left column, we show the energy-strain curves for the three polymer (FJC) models: the PDA model $(\diamondsuit)$, the PNA model $(\vartriangle)$ and the PAB model $(\star)$. The inset shows the three Hookean models: the HDA model $(\triangledown)$, the HNA model $(\circ)$ and the HAB model $(\square)$. The right column shows nominal stress-strain curves for the PDA $(\diamondsuit)$, the PNA $(\vartriangle)$, the PAB $(\star)$, the HDA $(\triangledown)$, the HNA $(\circ)$ and HAB models $(\square)$. All models are evaluated at $8N=400$, $b=0.1$, $a_0=1$, $\kt=1$.}
 \end{figure*}

\begin{figure}
\includegraphics[width=\linewidth]{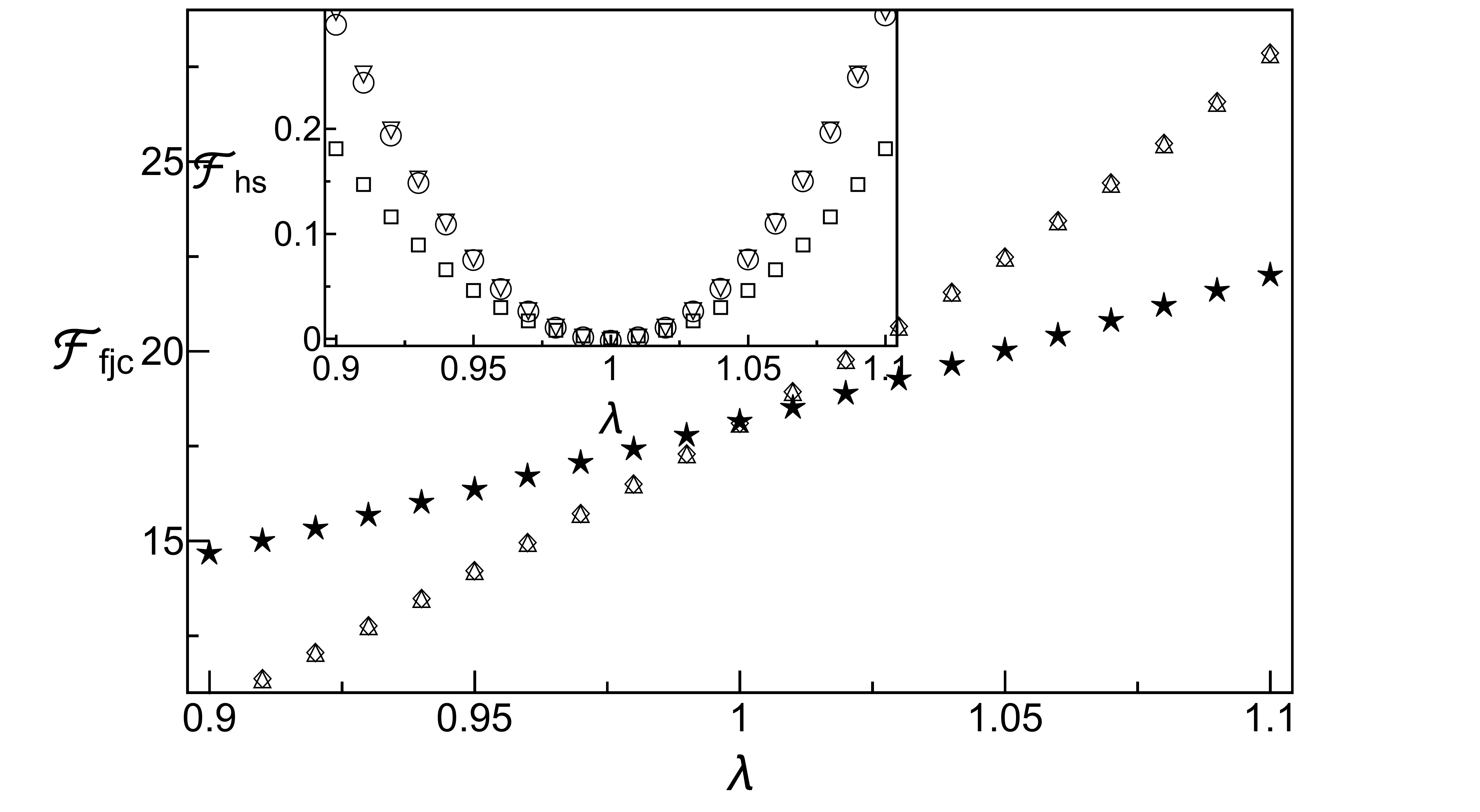}
\includegraphics[width=\linewidth]{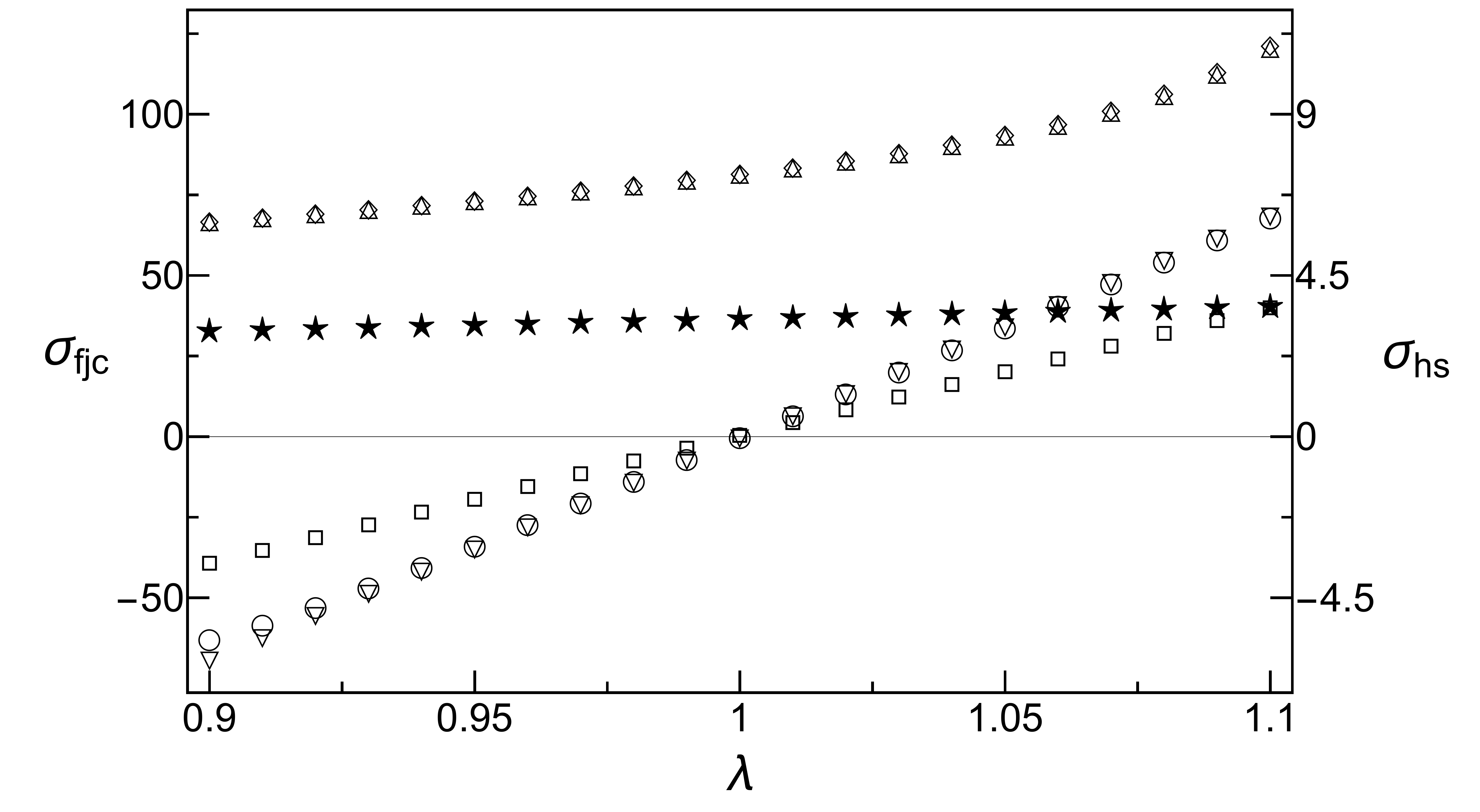}
\caption{\label{eb}Energies (upper panel) and stresses (lower panel) versus applied strain $\lambda$ in {\bf uniform extension} deformation $\sla_U$. In the upper panel, we show the energy-strain curves for the three polymer (FJC) models: the PDA model $(\diamondsuit)$, the PNA model $(\vartriangle)$ and the PAB model $(\star)$. The upper panel inset shows the three Hookean models: the HDA model $(\triangledown)$, the HNA model $(\circ)$ and the HAB model $(\square)$. The lower panel shows nominal stress-strain curves for the PDA $(\diamondsuit)$, the PNA $(\vartriangle)$, the PAB $(\star)$, the HDA $(\triangledown)$, the HNA $(\circ)$ and HAB models $(\square)$. All models are evaluated at $8N=400$, $b=0.1$, $a_0=1$, $\kt=1$.}
\end{figure}

Each of the six models is subjected to the five different deformations. We plot the measured energies, and the measured stresses - evaluated at precisely the same chain densities and, for the distributed linker models, averaged over the same ensemble of linker positions, in Figures \ref{eua}-\ref{eb}. Plotted are the total energies
\be
\cF_{\rm tot}=\sum_{i=1}^8 \cF(N_i,\ell_i,T)\, ,
\ee 
where $\cF(N_i,\ell_i,T)$, the energy contribution of each chain separately, is given either by Eq. \ref{kspH} or Eq. \ref{enerH2}, depending on the chain elastic response, and the nominal stress (setting $p=0$),
\be
\sigma=\frac{1}{V}\frac{\dd \cF_{\rm tot}}{\dd \lambda}\, .
\ee
We consider one deformation type where volume of the unit cell is not preserved - the uniform expansion/compression, where the stress needs to be corrected. 
We use of this infinitesimal deformation mode to establish the pre-stress. Note the symmetries in the energy-strain curves (upper panel of Fig. \ref{eb}): PDA, PNA and PAB energies are odd around the ground state ($\lambda=1$), unlike HDA, HNA, HAB that are even around the same state. In the lower panel of the same Fig. \ref{eb} stress-strain curves of the same pre-stressed models, PDA, PNA and PAB show finite stresses at no deformation ($\lambda=1$). A lot of information is presented in rather compact format in these graphs but we distill from it three key observations.
\begin{itemize}
\item{\bf Observation 1:} In uniform extension, FJC models have nonzero stress at zero strain while Hookean models do not.
\item{\bf Observation 2:} For all cases considered, the PNA and PDA models give indistinguishable results. For FJC models, disorder matters but affinity/non-affinity does not.
\item{\bf Observation 3:} Both classes considered throughout this paper (FJC and Hookean models) show the same type of behavior: introducing disorder leads to a more rigid mechanical response when compared to Arruda-Boyce settings.
\end{itemize}

{\bf Re: Observation 1:} A finite tension is required to extend an FJC from a corner of the unit cell to the linker, and although the vector sum of these tensions is zero the average tension is not zero, nor is it isotropic at the length scale of this unit cell. 
Were we to allow the boundary to relax, the unit cube would collapse to zero which is why the PAB model, analogous to what happens in classical rubber elasticity theory \cite{Miehe2011}, must be stabilized in the ground state by a uniform hydrostatic pressure (pre-stress) to ensure that the cube with side $a_0$ is indeed the equilibrium volume. 
The value of this uniform pre-stress for the PAB model may be obtained by applying no deformation to the system (i.e., setting all $\lambda_i$ equal to $1$ in Eq. \ref{sigc}).
\be
\sigma_{PS}=\frac13 \sqrt{N} \InvLgv\left(\frac{1}{\sqrt{N}}\right)\,.
\ee
For the PNA and PDA it is higher, because the distributed $N_i$ yield higher contributions to the average of $\InvLgv\left(\frac{1}{\sqrt{N}}\right)$ from the smaller $N_i$ than from the larger ones ($\InvLgv(x)$ rises steeply with $x$). 
We have defined the Hookean spring models such that there is no pre-stress in the reference configuration by setting each spring's equilibrium length to its value in the undeformed state, for given linker placement.  While this observation is quite obvious, the pre-stresses have important consequences for the non-affinity.

{\bf Re: Observation 2:} Surprisingly, the PNA model and PDA model give very comparable results. For FJC unit cells, the non-affinity appears negligible. 
We verify that this is the case in Sec. \ref{sec5}. 
This is not a universal feature: the equivalently defined Hookean models do, generally, show a strong effect of non-affinity which, except for in uniform extension, makes the material softer in all deformations considered. 
This softening by non-affinity is a general feature, which may be attributed to the number of constraints on the mechanical system: the non-affine system is less constrained and therefore better able to accommodate strains by rotating (which costs no elastic energy) rather than stretching. 
We propose that the suppression of non-affinity in the FJC models is a direct result of the pre-stress, which acts to effectively constrain the otherwise freely moving linker. 
This same effect, in extremes, may be seen in so-called marginal materials which may be completely floppy until they are stabilized by external or internal stresses \cite{DennisonStorm}. If this is the case, then we should be able to suppress the non-affinity by externally applying a pre-stress to these unit cells as well - we demonstrate this effect in Sec. \ref{sec5}.

{\bf Re: Observation 3:} The first part - that disordered FJC models are stiffer than the center-linker Arruda-Boyce model, 
may be understood by noting that redistributing a fixed amount of chain contour length in a unit cell of fixed size necessarily yields chains shorter as well as longer than the average length (rather than eight chains of equal length). 
The lack of non-affine displacements, noted in Observation 2, implies that the springs inside the unit cube deform uniformly and act in parallel. The presence of non-affine displacements in ensembles of springs that act in series has been noted before \cite{DiDonnaLubensky2005}.
The stiffness of entropic chains scales inversely with the contour length, the shorter (stiffer) chains dominate the average rigidity. Each of the eight springs has a spring constant
\be
k_i=\frac{3 \kt}{N_i b^2}\sim \frac{1}{N_i}
\ee
Comparing the two average spring constants, i.e. $\frac{1}{8} \sum_i k_i$ versus  $\frac{1}{8} \sum_i k_{AB}$, after discarding constant prefactors and rewriting the number of segments per AB chain as $\Nab=\frac{1}{8}\sum_i\Nseg$, we may write
which holds for all partitions of $N_{\rm tot}=\sum_i N_i$. Thus, the average and effective spring constant in the distributed system is always greater than it is for the centrally placed linker.

In Tables II and III we list the effective linear moduli extracted from our simulations. These are obtained by fitting the energy graphs in Figs. \ref{eua}-\ref{eb} to a harmonic function
\be
{\cal E}(\lambda)\approx \frac12\, \chi\, (\lambda-\lambda_{\rm eq})^2\, ,
\ee
where $\chi$ may be a shear or a bulk modulus, depending on the deformation considered, and $\lambda_{\rm eq}$ is the $\lambda$ corresponding to the underformed state, {\em i.e.} $\lambda_{\rm eq}=1$ for uniaxial, biaxial, pure shear and uniform extension, and $\lambda_{\rm eq}=0$ for simple shear. 
Table II shows that in the Hookean models, positional disorder (second column) tends to increase all linear stiffnesses compared to the symmetric AB model (first column), and that non-affinity (third column) softens the disordered system considerably. 
In Appendix \ref{App}, we demonstrate how the $\chi$'s for the HAB model can be determined analytically to validate our method. Interestingly, the uniform extension deformation is itself symmetric, and absent elastic nonlinearities (as they are in the linearized regime we consider here) will be absorbed affinely. This explains the absence of the typical softening of the uniform extension modulus between affine and non-affine disordered models.

The differences between the FJC models and the Hookean models are in the amounts of pre-stress. The FJC models are unavoidably pre-stressed, while the Hookean models are not. 
These pre-stresses, we argue, serve to constrain otherwise free degrees of freedom within the system, and are thus directly related to the disappearance of the non-affinity in the FJC models. We now explore this effect by applying external pre-stresses to Hookean systems.

 \begin{table}
 \caption{Linear moduli $\chi$ for Hookean spring models} \label{defhs}
 \centering
 \begin{tabular}{c c c c}
 Deformation & $\chi_{\rm HAB}$ & $\chi_{\rm HDA}$ & $\chi_{\rm HNA}$\\
 \hline
 Uniform extension & 36 & 62.616 & 61.677 \\
 Pure shear & 0 & 5.809 & 0.453 \\
 Uniaxial & 0 & 4.34 & 0.34  \\
 Biaxial & 0 & 17.396 & 1.38199\\
 Simple shear & 4 & 5.996 & 5.245  
 \end{tabular}
 \end{table}

\begin{table}
 \caption{Linear moduli $\chi$ for FJC models} \label{deffjc2}
 \centering
 \begin{tabular}{c c c c}
 Deformation & $\chi_{\rm PAB}$ & $\chi_{\rm PDA}$ & $\chi_{\rm PNA}$\\
 \hline
 Uniform extension & 38 & 191.128 & 188.23 \\
 Pure shear & 48.881 & 116.793 & 116.095 \\
 Uniaxial & 36.663 & 87.489 & 86.929  \\
 Biaxial & 146.634 & 349.889 & 348.426\\
 Simple shear & 12.3748 & 38.321 & 37.605  
 \end{tabular}
 \end{table}
 
\section{Pre-stress suppresses non-affinity} \label{sec5} 

\begin{figure*}
\subfloat[uniaxial deformation]{\includegraphics[width = 7cm]{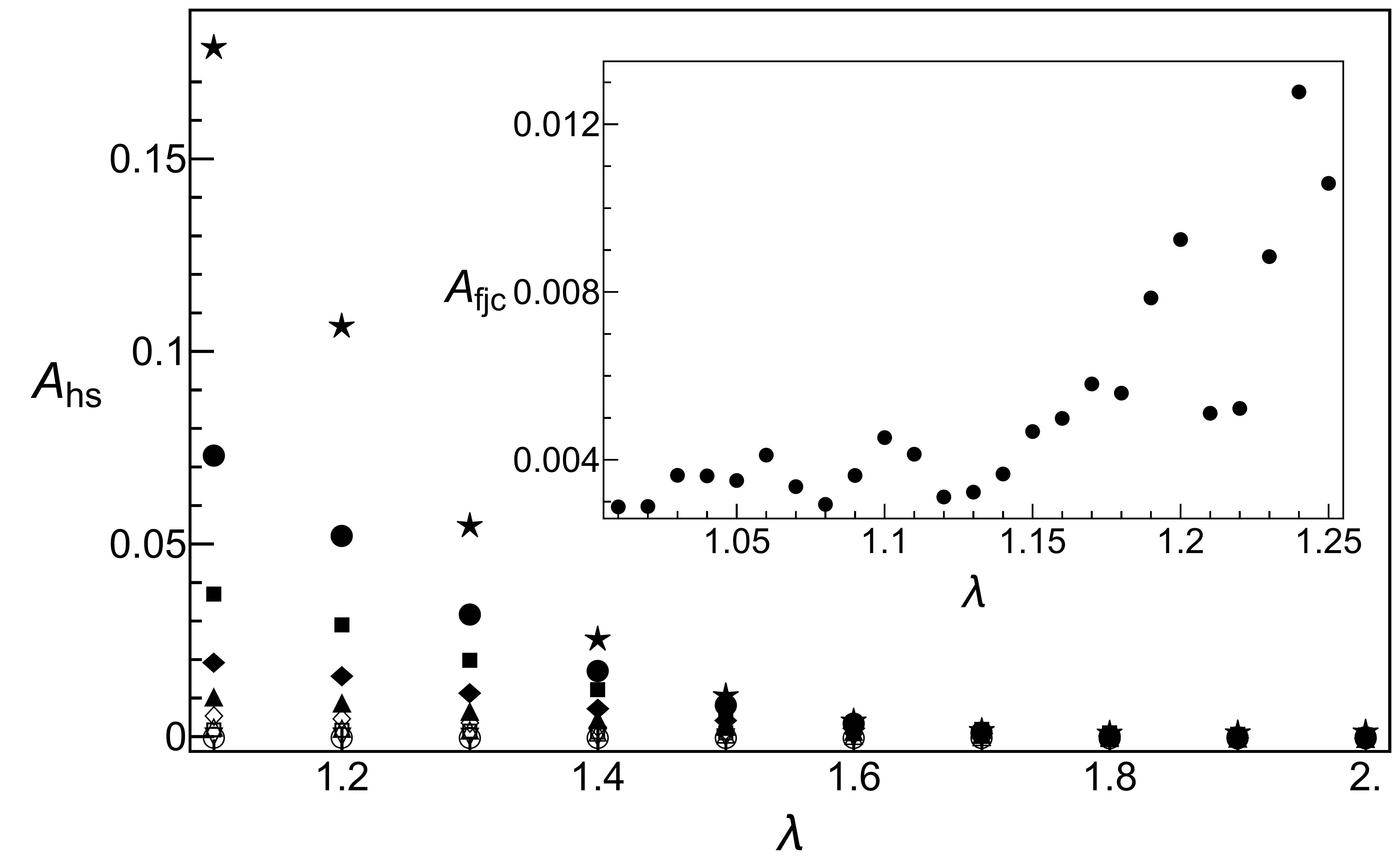}}
\subfloat[biaxial deformation]{\includegraphics[width = 7cm]{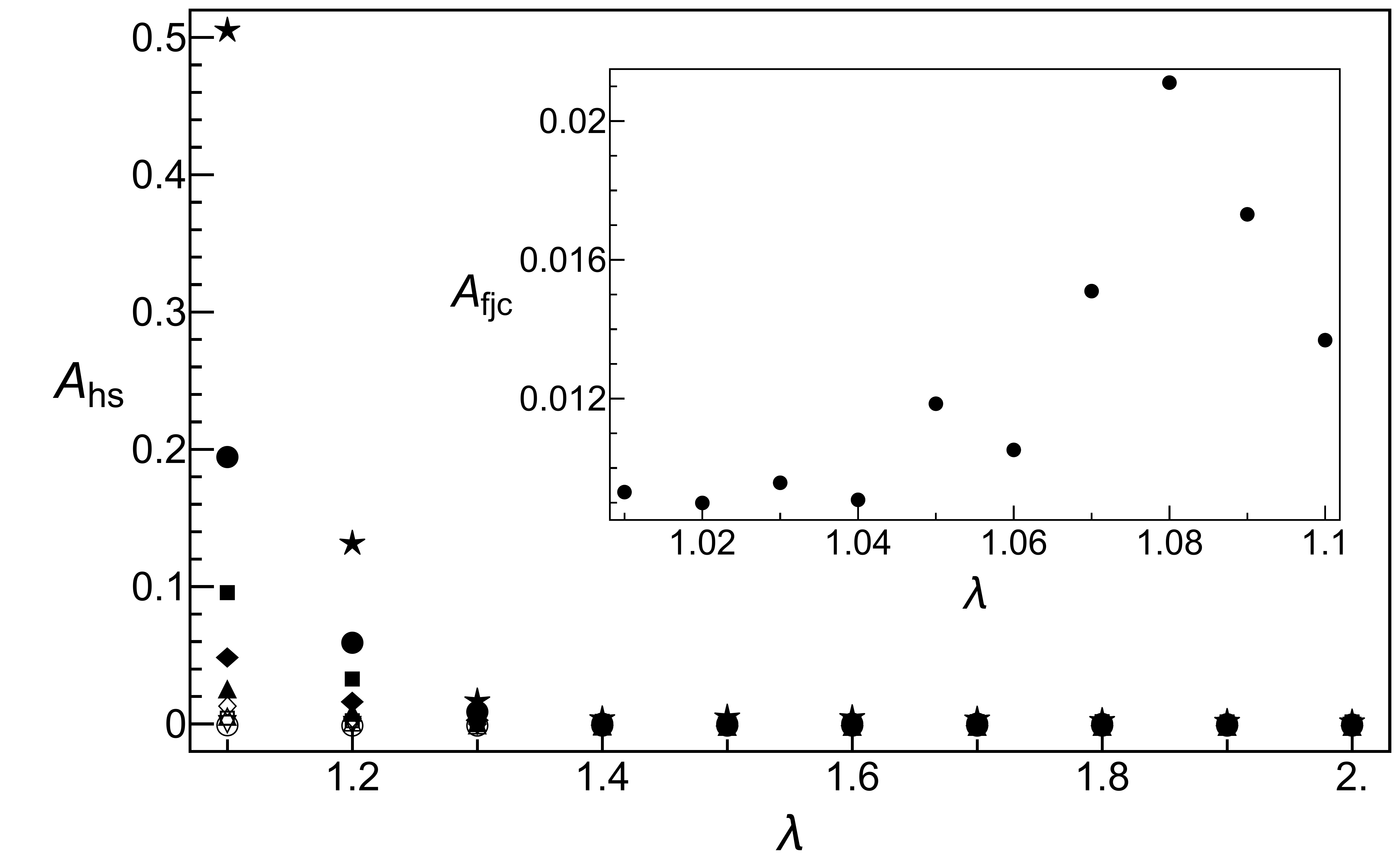}}

\subfloat[pure shear]{\includegraphics[width = 7cm]{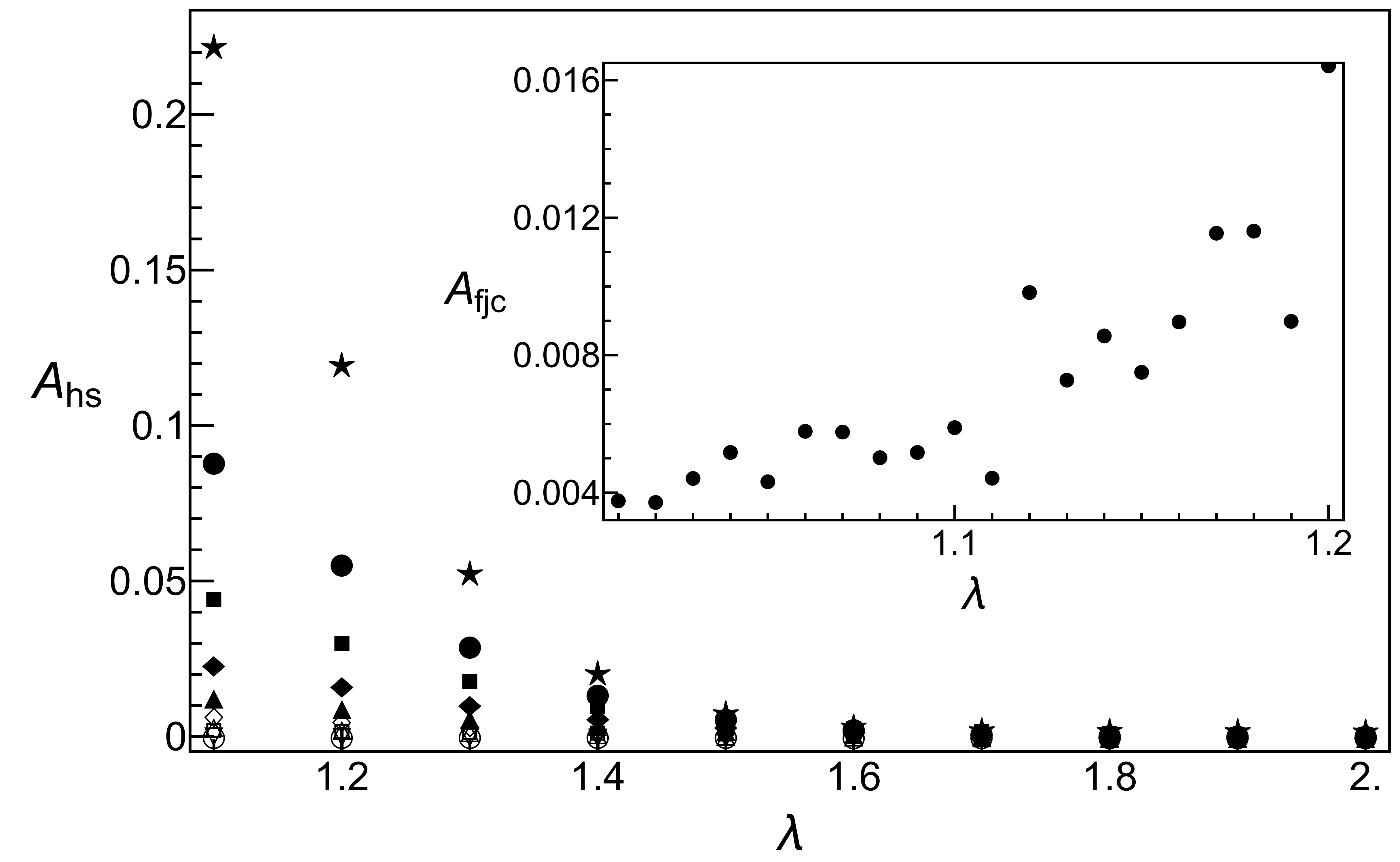}}
\subfloat[uniform extension/compression]{\includegraphics[width = 7cm]{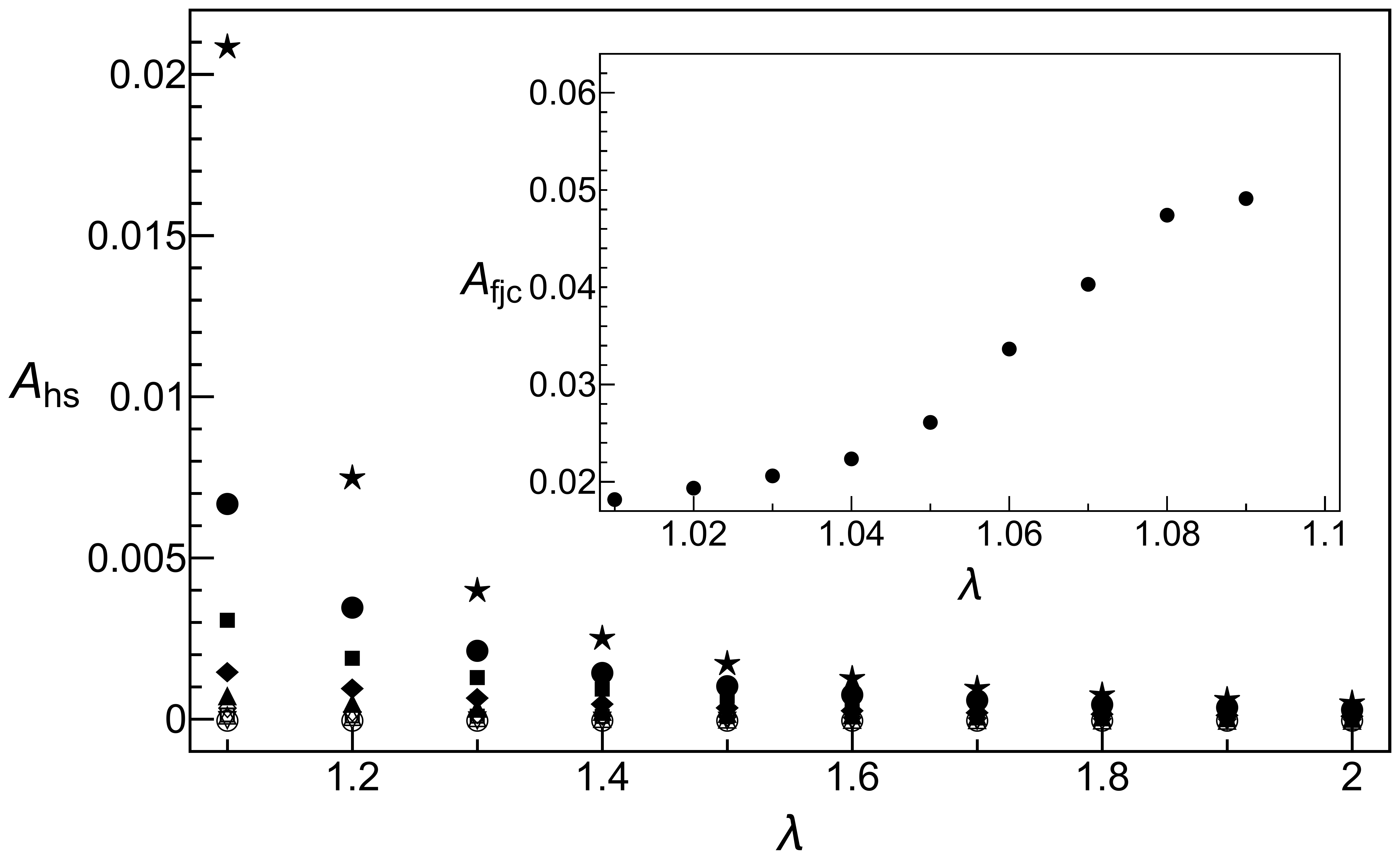}}

\caption{\label{nam}Ensemble-averaged differential non-affine measure $\mathbf{A}_{\textrm{hs}}$ versus strain $\lambda$ for stress-free and pre-stressed HNA models for the applied deformation. In the main figures, the initial (zero-strain) lengths of individual filaments are set to $100\%(\star)$, $90\%(\bullet)$, $80\%(\blacksquare)$, $70\%(\blacklozenge$), $60\%(\blacktriangle)$, $50\%(\diamondsuit)$, $40\%(\vartriangle)$, $30\%(\square)$, $20\%(\triangledown)$, $10\%(\circ)$ of the corner-to-linker distance. The inset graphs show the ensemble-averaged differential non-affinity measure $\mathbf{A}_{\textrm{fjc}}$ versus strain for the PNA model, as defined in the text.}
\end{figure*}

\begin{figure}
\includegraphics[width=\linewidth]{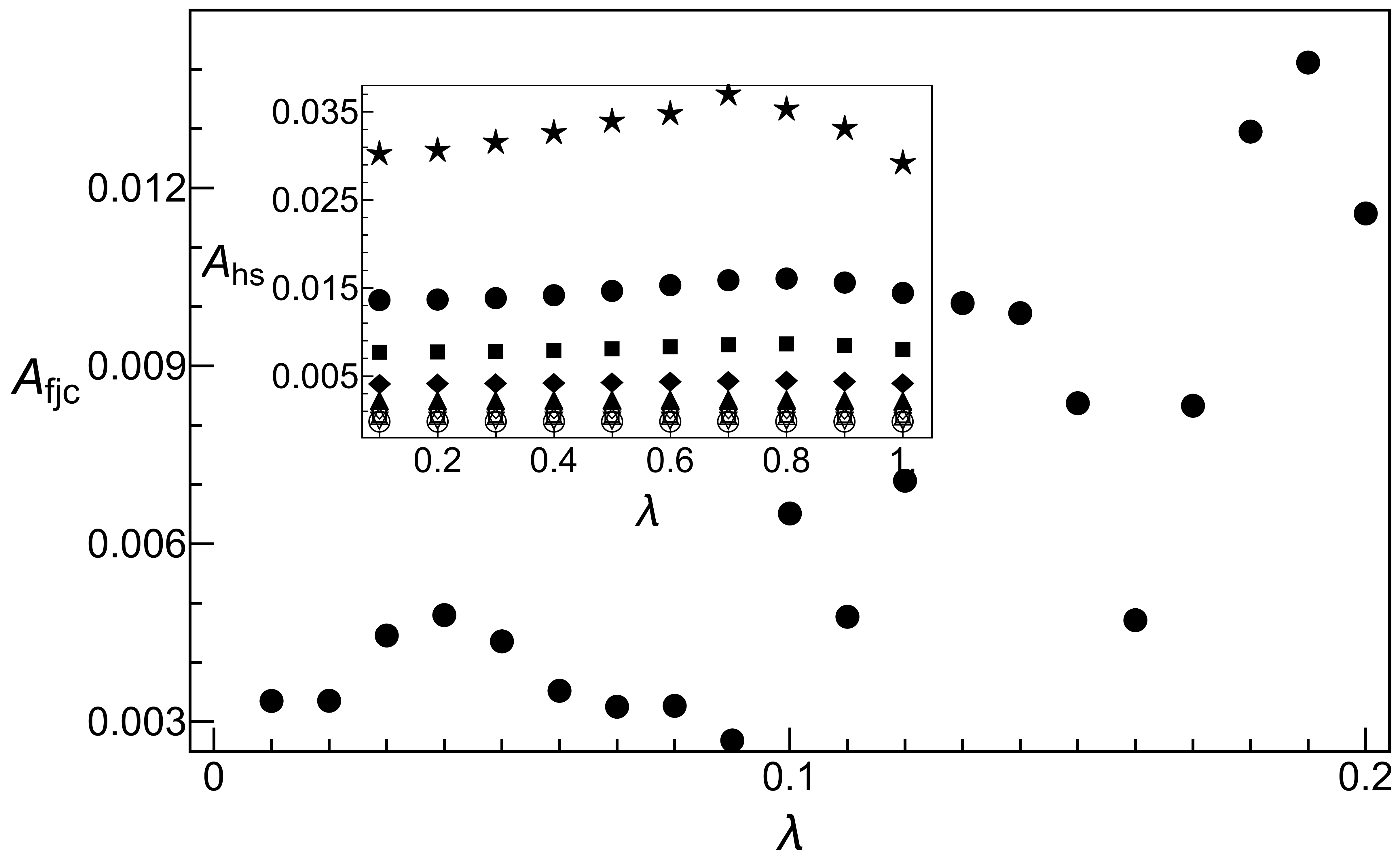}
\caption{\label{namss}Ensemble-averaged differential non-affine measure $\mathbf{A}_{\textrm{hs}}$ versus strain $\lambda$ for stress-free and pre-stressed HNA models for {\bf simple shear} deformation. The main figure shows the ensemble-averaged differential non-affinity measure $\mathbf{A}_{\textrm{fjc}}$ versus strain for the PNA model, as defined in the text. In the main figure, the initial (zero-strain) lengths of individual filaments are set to$100\%(\star)$, $90\%(\bullet)$, $80\%(\blacksquare)$, $70\%(\blacklozenge$), $60\%(\blacktriangle)$, $50\%(\diamondsuit)$, $40\%(\vartriangle)$, $30\%(\square)$, $20\%(\triangledown)$, $10\%(\circ)$ of the corner-linker distance.}
\end{figure}

To directly measure the effect of pre-stress on non-affinity, we proceed now by measuring the amount of non-affinity as a function of strain, for differently pre-stressed systems. 
Several methods of quantifying the non-affinity have been proposed \cite{DiDonnaLubensky2005, HeadLevine2003, FalkLanger1998, HuismanDynamical, Hatami-Marbini2008, Hatami-Marbini2009}. 
For our systems we will use the differential measure proposed by Huisman et al. \cite{HuismanDynamical} and effectively measure local deviations from an affine deformation field at every strain step. 
As remarked in the previous, models with a centrally placed linker (PAB and HAB) will be affine by symmetry, and disordered models where we force affine deformations (PDA, HDA)) will, obviously, behave affinely as well. 
Thus we compare, in this section, the response of the two non-affine models, PNA and HNA. PNA, as we have seen, is intrinsically pre-stressed and shows little non-affinity. 
HNA is not pre-stressed and shows non-affine softening in all moduli that we have measured.  

Because we consider a range of deformations, we require a slightly more general differential non-affinity measure than \cite{HuismanDynamical}. 
Consider, to this end, a linker that at some strain value $\lambda$ is located at $\mathbf{X}_{\lambda}$. Where will it be, after a strain increment $\delta \lambda$, if it moves affinely from its previous position? 
We define the incremental strain tensor $\tilde \sla$, as follows
\be
\sla(\lambda+\delta \lambda)\equiv \tilde \sla(\lambda;\delta \lambda)\cdot \sla(\lambda)\, .
\ee
By expanding the strain tensor around $\lambda$, we obtain that
\be
\tilde \sla(\lambda;\delta \lambda)=\mathbb{1}+\left(\left.\frac{\partial\sla}{\partial\lambda}\right|_\lambda \cdot\sla^{-1}(\lambda)\right)\delta \lambda
\ee
For a simple shear deformation, $\tilde \sla_S(\lambda;\delta \lambda)$ is simply $\sla_S(\delta \lambda)$, independent of the current state of strain. For more general deformations, however, it will depend on $\lambda$ - for instance, $\tilde \sla_V(\lambda;\delta \lambda)$ equals $\sla_V(1+\delta \lambda/\lambda)$. 

We construct the differential measure of non-affinity by squaring the non-affine deviation - {\em i.e.}, the actual position $\mathbf{X}_{\lambda+\delta\lambda}$ minus the affinely propagated position $\tilde \sla(\lambda;\delta \lambda)\cdot\mathbf{X}_{\lambda}$ - incurred during the strain step $\delta \lambda$, normalized to the magnitude of the strain step:
\be
\mathbf{A}=\frac{1}{(\delta \lambda)^{2}}\left|\mathbf{X}_{\lambda+\delta\lambda}-\tilde{\sla}\left(\lambda;\delta\lambda\right)\cdot \mathbf{X}_{\lambda}\right|^{2}
\ee
This quantity is plotted against the strain $\lambda$, for all deformation modes considered, for the two non-affine models (PNA and HNA) in Figs. \ref{nam}-\ref{namss}. 
For the HNA models, we apply an external pre-stress by reducing the rest lengths of all springs by a uniform factor. The resulting network will generally require a second relaxation towards a new ground state. 
We strain the systems from this new, equilibrated, configuration to obtain the measures displayed here. All HNA graphs clearly demonstrate that indeed, also in the Hookean models pre-stress results in diminished non-affinity. 
What these graphs also show, is that although the non-affinity is small for the FJC models, it is not zero. 
Notably, even the non-prestressed Hookean systems show a continued downward trend in differential non-affinity at finite strain, whereas the FJC systems in all cases show a rise over the regime considered. 
The differential measure is expected to tend to zero in the large strain limit. At the highest forces, the linker is completely constrained and no longer free nor able to move other than affinely. 
This high strain limit, however, may not be attained in the FJC models as these are intrinsically finitely extensible: it is impossible to extend them beyond a contour length of $N b$. 
We postulate that in our systems, the chains are too short to attain the affine limiting behavior. Our Hookean systems show a decreasing non-affinity, while the FJC models show a small but rising non-affinity. 
While there are very few studies that measure the strain-dependence of the non-affinity, both behaviors have actually been reported. 
The increase of non-affininty is seen in simulations on Worm Like Chain systems, which also have pronounced non-linear elasticity but not, necessarily, pre-stresses. 
Fibrin gels, in the only experimental measurement that we are aware of, display a pronounced peak in non-affinity before entering a prolonged regime of diminishing non-affine deformations \cite{WenBasu}. 
We see, that depending on the deformation mode considered the amount of non-affinity varies greatly - it appears that some deformations are more conducive to non-affinity than others.

\section{Summary and Conclusions} \label{sec6}

We set out to assess what the effect of spatial and elastic disorder, single-chain elasticity and pre-stress is on the mechanical response of simple unit cell network models for rubber-like elasticity. 
Models based on the Freely Jointed Chain are invariably pre-stressed. 
Addition of a uniform pressure stabilizes a state, but does not remove pre-stress. FJC models are intrinsically less non-affine than Hookean spring models, even when these are, otherwise, elastically equivalent. 
The effects of externally applied pre-stress in Hookean systems are similar, as they too suppress the amount of non-affinity, but closer inspection reveals that the manner in which non-affinity depends on strain is still quite different. We speculate that this is due to the nonlinear force extension relation of the FJC.

In reality, many materials are indeed pre-stressed. For biopolymer gels and tissues, in particular, the presence of pre-stresses is well-documented. 
Previous modeling \cite{Storm2005} assumed complete affinity, yet was able to capture the mechanical response of a variety of biopolymer gels reasonably well. A possible explanation for the success of such affine models, we propose, may be that non-linearly elastic polymer models, even in the most basic incarnations considered here, display a greatly suppressed non affinity and are, by many accounts, very well approximated by fully affine models.

This finding also suggests one should proceed with caution in replacing nonlinear constituents with linear springs. If one removes the pre-stress, this may result in an unrealistically high degree of non-affinity and, accordingly, lower network stiffness. Even if one linearizes the springs retaining an appropriate amount of pre-stress, the non-affine response may still evolve very differently with strain due to the nonlinearities.

In those experiments that are able to map out the non-affinity of the crosslinkers (or of otherwise fixed fiducial markers inside the network) \cite{DiDonnaLubensky2005,FalkLanger1998,WenBasu,LiuKoenderink,Gardel2003}, it should be feasible to investigate directly the correlation between pre-stress and non-affinity. We predict these two to be inversely related.

In upcoming work, we explore the effects of polymer persistence on the effects reported here. The associated Worm-Like Chain model features a force-extension relation that has both a finite rest length, as well as a steeply nonlinear response to stretching. As such, semiflexible chains represent an interesting middle ground between the Hookean and FJC cases considered here. It is entirely unclear which effect will be dominant in their mechanical response: the lack of ground state pre-stresses promoting non-affinity, or the nonlinear force-extension that suppresses it.

\section{Acknowledgments}
We thank F.C. MacKintosh, A. Sharma, H.E. Amuasi W.G. Ellenbroek, F. Fontenele Araujo and M.A.J. Michels for valuable discussions. This work was generously supported by the Institute for Complex Molecular Systems (ICMS) at the Eindhoven University of Technology, and is also part of the Industrial Partnership Programme (IPP) Bio(-Related) Materials (BRM) of the Stichting voor Fundamenteel Onderzoek der Materie (FOM), which is supported financially by Nederlandse Organisatie voor Wetenschappelijk Onderzoek (NWO). The IPP BRM is
cofinanced by the Top Institute Food and Nutrition and the Dutch Polymer Institute.

\section{Bibliography}

\bibliographystyle{elsarticle-num} 
\bibliography{references}

\appendix \label{App}
\section{Numerical computation of moduli and exact expressions for the HAB model}
As detailed in the text, we compute moduli by determining, numerically, the curvature of the energy vs. strain graph. In one case considered here, we can compute this exactly. 
For the HAB model, we have eight chains that each contribute identically to the strain energy. One of them, in the undeformed state, has as its end-to-end vector $v=(a_0/2)(\hat x + \hat y +\hat z)$, and its length is $\sqrt{3 a_0}/2$ \cite{Boal}. Acting on this polymer with, for instance, the deformation tensor for simple shear, it is transformed affinely to $v'=\sla_S(\lambda)\cdot v$. The length of this deformed vector is 
\be
|v'|=\frac12 a_0 \sqrt{\lambda^2+2\lambda+3}\, .
\ee
Writing the deformation energy of each chains as $(1/2) k_{\rm sp} \delta \ell^2$, multiplying by 8 for the number of chains in the unit cell, and dividing by the unit cell volume we arrive at the strain energy density for the HAB in simple shear
\be
{\cal E}_S(\lambda)=\frac{k_{\rm sp}}{a_0}\left(\sqrt{\lambda^2+2\lambda+3}-\sqrt{3}\right)^2\, .
\ee 
To extract the modulus, we expand to second order in $\lambda$:
\be
{\cal E}_S(\lambda)\approx \frac12 \left(\frac{2 k_{\rm sp}}{3 a_0}\right)\lambda^2+{\cal O}(\lambda^3)\, ,
\ee
From which we read off the modulus in simple shear
\be
\chi_S=\frac{2 k_{\rm sp}}{3 a_0}.
\ee
For the values we have chosen in the simulations ($N=50, b=0.1, \kt=1, a_0=1$, the spring constant computed with Eq. \ref{kspH} is $k_{\rm sp}=6$ and thus $\chi=4$. This was confirmed in the simulations. All other moduli may be similarly computed - it turns out that the only other nonzero one is the bulk modulus probed in uniform extension; $\chi_V=(6k_{\rm sp}/a_0)=36$.

\end{document}